\documentstyle[preprint,aps,array,epsf]{revtex}

\newcommand{\beq}[1]{\begin{equation} \label{(#1)}}
\newcommand{\eeq}{\end{equation}}
\newcommand{\ba}[1]{\begin{eqnarray} \label{(#1)}}
\newcommand{\ea}{\end{eqnarray}}
\newcommand{\nn}{\nonumber}
\newcommand{\rf}[1]{(\ref{(#1)})}

\def\lsim{\mathrel{\vcenter{\hbox{$<$}\nointerlineskip\hbox{$\sim$}}}}
\def\gsim{\mathrel{\vcenter{\hbox{$>$}\nointerlineskip\hbox{$\sim$}}}}

\def\th{$^{\rm th}$}
\def\half{\frac{1}{2}}

\def\th{^{th}}
\def\nusun{\nu_\odot}
\def\nue{\nu_e}
\def\nunote{\nu_{\not e}}
\def\numu{\nu_{\mu}}
\def\nunotmu{\nu_{\not \mu}}
\def\nut{\nu_\tau}
\def\nutau{\nu_\tau}
\def\nus{\nu_s}
\def\num{\nu_-}
\def\nup{\nu_+}

\def\numubar{\bar{\nu}_{\mu}}

\def\nuH{\nu_{\rm H}}
\def\dmsq{\delta m^2}
\def\dmatm{\delta m^2_{\rm atm}}
\def\dmsun{\delta m^2_{\rm sun}}
\def\dmlsnd{\delta m^2_{\rm LSND}}
\def\thetam{\theta_M}
\def\thetav{\theta_V}

\def\lambdav{\lambda_V}
\def\notmu{\not\mu}
\def\epsee{\epsilon_{ee}}
\def\epsmm{\epsilon_{\mu\mu}}
\def\epsme{\epsilon_{\mu e}}
\def\thetasun{\theta_{\rm sun}}
\def\thetaatm{\theta_{\rm atm}}
\def\thetats{\theta_{\tau s}}
\def\eps{\epsilon}
\def\Ratm{R_{\rm atm}}
\def\Rsun{R_{\rm sun}}

\begin{document}
\title{The Hidden Sterile Neutrino and the (2+2) Sum Rule}
\author{Heinrich P\"as,$^1$
\footnote[1]{E-mail: paes@physik.uni-wuerzburg.de}
Liguo Song,$^2$
\footnote[2]{E-mail: liguo.song@vanderbilt.edu}
and Thomas J. Weiler$^2$
\footnote[3]{E-mail: tom.weiler@vanderbilt.edu}
}
\address{$^1$ Institut f\"ur Theoretische Physik und Astrophysik, 
Universit\"at W\"urzburg,\\ Am Hubland, 97074 W\"urzburg, Germany }
\address{$^2$ Department of Physics and Astronomy,
Vanderbilt University\\
Nashville, TN 37235, USA}

\maketitle

\begin{abstract}
We discuss oscillations of atmospheric and solar neutrinos 
into sterile neutrinos in the 2+2 scheme. 
A zero$^{th}$ order sum rule requires 
equal probabilities for
oscillation into $\nu_s$ and $\nut$ 
in the solar+atmospheric data sample.
Data does not favor this claim.
Here we use scatter plots to assess corrections of the 
zero$^{th}$ order sum rule 
when (i) the $4\times 4$ neutrino mixing matrix assumes 
its full range of allowed values,
and (ii) matter effects are included.
We also introduce a related ``product rule''.
We find that the sum rule is significantly relaxed,
due to both the inclusion of the small mixing angles
(which provide a short-baseline contribution) and to matter effects.
The product rule is also dramatically altered.
The observed relaxation of the sum rule 
weakens the case against the 2+2 model 
and the sterile neutrino.
To invalidate the 2+2 model, a
global fit to data {\sl with the small mixing angles included}
seems to be required.
\end{abstract} 

\newpage

\section{Introduction}
Taken at face value, the solar, atmospheric, and LSND data 
require three independent $\dmsq$ scales.  
Thus, four neutrinos seem to be required.
The $Z$-boson width further requires that one of these
four neutrinos be a ``sterile'' $SU(2)\times U(1)$ electroweak-singlet.
Given the profound theoretical implication which would accompany 
the validation of a light sterile neutrino,
and given the present experimental search for 
the sterile neutrino by the mini-BooNE 
experiment \cite{miniBooNE}, 
a serious look at the present viability of the 
sterile neutrino is well motivated.

It was shown some time ago 
that the non-observation of $\nue$ and $\numu$
disappearance by the Bugey and CDHS experiments,
in conjunction with the appearance of $\numu \rightarrow \nue$
in the LSND experiment,
strongly disfavored the 3+1 spectral 
hierarchy \cite{BGG,Barger:1998bn} in which 
the large ``LSND'' mass-splitting isolates one heavier mass eigenstate
$|\nuH \!\!>$ from the other three.
The conflict results because the Bugey and CDHS rates are proportional
to $|<\nue|\nuH>|^2$ and $|<\numu|\nuH>|^2$, respectively,
whereas the LSND rate is proportional to the product of the two;
the upper limits from Bugey and CDHS doubly-suppress the rate 
allowed for LSND.
However, when the final value of the LSND 
oscillation amplitude was reported to be smaller 
\cite{Mills:2001tq} than earlier results, 
the 3+1 spectrum was resurrected at the 95\% confidence limit 
for four isolated values of 
$\dmsq$ \cite{Barger:2000ch,Smirnov:2000pa,Giunti:2001ur}.
Recently, 
a careful analysis of all data has led to the conclusion that even these
isolated mass values are ``highly unlikely,''
viable only if one super-conservatively 
accepts the tiny intersections of the  99\% Bugey/CDHS exclusion curve 
with the 99\% allowed region of LSND \cite{Grimus:2001mn}.
This 3+1 spectrum is a ``separated-sterile'' model,
in that the sterile flavor resides dominantly in the isolated 
mass-state and barely mixes with the three active flavors;
in the absence of the LSND data, the active-sterile mixing can be 
made arbitrarily small.


On the other hand, the alternative 2+2 mass spectrum is compatible
with the combination of short baseline data.
In the 2+2 spectrum, the LSND scale splits
two pairs of neutrino mass-eigenstates.
Phenomenologically, it is required that one pair mix $\numu$ with
$\nut$ and $\nus$ to explain the atmospheric $\numu$ disappearance,
while the second pair mix $\nue$ with $\nut$ and $\nus$
to explain the solar $\nue$ disappearance.  
The small LSND amplitude, and the Bugey and CDHS limits are 
accommodated with a small mixing of $\nue$ into the first pair,
and small mixing of $\numu$ into the second pair \cite{Barger:1998bn}.
In this model, the sterile neutrino is distributed in its totality 
between the solar and atmospheric scales.
This leads to an approximate sum rule \cite{Peres:2001ic}.
The essence of the sum rule is that the sterile neutrino
may hide from solar oscillations, or from atmospheric oscillations,
but cannot hide from both.
When matter effects and small angles mixing the atmospheric 
and solar pairs are neglected, the sum rule states that 
the probabilities to produce $\nus$ in the solar and atmospheric 
data sum to unity.  

The Superkamiokande (SK) experiment has put a limit on the
atmospheric mode $\numu \rightarrow \nus$ \cite{newSKlimit}.
In principle, the atmospheric data is sensitive to sterile neutrinos 
in three ways: 
(i) through differing matter effects for $\numu$ oscillation to 
$\nut$ vs.\ $\nus$;
(ii) through neutral current scattering of $\nut$ but not $\nus$, 
as measured by $\pi^0$ production;
and (iii) through $\tau$ appearance from $\nut$ scattering 
above threshold.
However, the SuperK analysis, following the guidelines in \cite{FLM01},
ignores all parameters but
$\dmatm$, $\thetaatm$, and the $\nutau$-$\nus$ mixing angle
$\thetats$. 
We will argue that the short baseline contribution 
from the large mass-gap $\dmlsnd$ is important if the 
small mixing angles are not too small,
in which case the SuperK analysis is invalid for the 2+2 model.

Limits on the solar mode $\nue \rightarrow \nus$ result from 
model fits to the SK solar data \cite{Giunti:2000wt}, but especially
from the recent SNO data.
There is no evidence favoring a sterile admixture in the 
neutrino flux from the sun.  
Nevertheless, a large sterile component 
remains compatible with the SNO result \cite{BMW01,snosterile}, 
because of our uncertainty in the true high-energy solar 
neutrino flux produced by the $^8$B reaction in the sun.
It is unclear whether future measurements of intermediate-energy neutrinos in 
the BOREXINO experiment or the solar neutrino component of the KAMLAND 
experiment will help resolve this issue.



Combined fits to solar and atmospheric 
neutrinos \cite{Gonzalez-Garcia:2001uy,Maltoni:2001bc,newValle} 
have been performed.
The authors of \cite{Maltoni:2001bc} came recently to the conclusion
that a global analysis considering short baseline, 
solar, and atmospheric
data gives a slightly better fit to (3+1) mass schemes compared to (2+2) 
schemes, though both four-neutrinos schemes are presently viable 
at low confidence.
Most recently, a global analysis \cite{newValle} 
has been performed which includes  
the LSND data, and newest SNO, SuperK, and MACRO data.
The conclusion reached is that both the 2+2 and 1+3 
sterile models are highly disfavored; 
the 2+2 analysis returns a goodness of fit of only $1.6\times 10^{-6}$.
However, to reduce the number of parameters involved, these papers set 
the two smallest mixing angles tested by short baseline experiments 
to zero in their fits to the solar and atmospheric data.

In this work, we analyze the approximate sum
rule in generality, varying the usually neglected mixing angles in
their experimentally allowed ranges, and including possible matter effects.
Our results indicate the allowed range of joint probability 
for $\nus$ to reside in the solar and atmospheric data.
We find that the allowed range may differ considerably from the 
zero$\th$ order result with small mixing-angles set to zero.
Although it is not known whether the same values of the small angles
which provide a large violation of the zero$\th$ order sum rule also 
provide good fits to the global data,
the sensitivity to small mixing-angles demonstrated here suggests 
that global fits which neglect the small angles may underestimate 
the viability of the 2+2 model.
This, in turn, casts some doubt on claimed exclusions of the 2+2 model.
The ultimate arbiter on the viability of the sterile neutrino will
be a global fit to data with all mixing angles included.\footnote
{Of the three small angles, 
we show that the sum rule is sensitive to $\epsmm$ and $\epsme$,
but not to $\epsee$ (in our notation).
$\epsmm$ has been included in recent global analyses.  Our findings 
suggest that, at a minimum, $\epsme$ should be included too.
}


\section{Formalism}
The starting point for neutrino oscillations 
is the unitary transformation between mass
and flavor basis:
\beq{U}
|\nu_\alpha> = \sum_j U^*_{\alpha j}\,|\nu_j> \,,
\quad\mbox{or}\quad <\nu_j\,|\,\nu_\alpha>=(U^\dag)_{j \alpha}\,;
\eeq
the inverse relation is 
\beq{Udag}
|\nu_j> = \sum_\alpha (U^\dag)^*_{j \alpha}\,|\nu_\alpha> \,,
\quad\mbox{or}\quad <\nu_\alpha\,|\,\nu_j>=U_{\alpha j}\,.
\eeq
We will consistently use roman indices for mass states,
and greek indices for flavor states.
The mass states $|\nu_j>$ and therefore the 
transformation matrix $U$ depend on matter densities,
and so we will often affix a superscript or subscript label 
$\kappa\in\{V,S,M,C\}$ to $|\nu_j>$ and $U$ 
to signify the vacuum, solar, earth's mantle, 
and earth's core environments, respectively.
In addition, the labels A and E will sometimes denote the 
atmosphere and the Earth.

The formula for the oscillation probability is
\beq{oscprob}
P^{\alpha\rightarrow\beta}=\delta_{\alpha\beta}-4\sum_{j<k}
\left[R^{\alpha j}_{\beta k}\,
\sin^2(\Phi_{kj})
+\half\, J^{\alpha j}_{\beta k}\,
\sin(2\Phi_{kj})
\right]\,,
\eeq
where the CP-conserving coefficient $R^{\alpha j}_{\beta k}$ 
and the CP-violating coefficient $J^{\alpha j}_{\beta k}$ 
are the real and imaginary parts, respectively, of 
$U_{\alpha j}U^*_{\beta j}U^*_{\alpha k}U_{\beta k}$,
and the relative phase $\Phi_{kj}\equiv\delta m^2_{kj}L/4E
=1.27\,(\delta m^2_{kj}/{\rm eV}^2)(L/{\rm km})/(E/{\rm GeV})$,
with $\delta m^2_{kj}=m^2_k -m^2_j$.
The oscillation length, determined by setting $\Phi_{kj}=\pi$,
is
\beq{osclength}
\lambda=2.48\,\left(\frac{{\rm eV}^2}{\delta m^2_{kj}}\right)
	     \left(\frac{E}{{\rm GeV}}\right)\,{\rm km} \,.
\eeq
This same probability formula holds in vacuum or in matter 
at near-constant density,
although the mixing elements and the effective masses differ
in these two situations.

After many oscillations, the coherence implied by the specific phases
in eq.\ \rf{oscprob} is lost, due to magnification of experimental 
uncertainties in $L/E$ or due to decoherence of the 
mass-eigenstate wave packets.  For either reason, the result is an 
averaging which takes $\sin^2 (\Phi_{kj})\rightarrow\half$ 
and $\sin (2\Phi_{kj})\rightarrow 0$.
This same averaging occurs also after even a few oscillations 
if the experimental uncertainties in $L/E$ are sufficiently large.
It is also important to note that the increase of $\lambda$ with $E$ 
may lead to the suppression of oscillations at high energy,
if the condition $\lambda \gg L$ is achieved. This suppression, 
together with matter effects, will become important in the discussion 
of through-going atmospheric neutrino events.

Four-flavor mixing is described by six angles and three CP-violating phases.
In addition there are three further phases for Majorana neutrinos;
since these three phases 
do not enter into oscillation probabilities, we omit them.
The six angles parametrize independent rotations in the
six planes of four-dimensional space.
For our purposes, it is useful to order these rotations as\footnote
{We note that the ordering in eq.\ \rf{Umatrix}
is the same as that in \cite{Gonzalez-Garcia:2001uy,Maltoni:2001bc,newValle}
when $\eps_{\mu e}$ and $\eps_{ee}$ are set to zero, 
as is done in \cite{Gonzalez-Garcia:2001uy,Maltoni:2001bc,newValle}.
}
%
%
\beq{Umatrix}
U =  R_{23}(\thetats) 
    R_{24}(\epsmm)  R_{14}(\epsme) R_{13}(\epsee)  
    R_{34}(\thetaatm) R_{12}(\thetasun) \,.
\eeq
%
Unitary $U$ transforms from the mass basis $(m_4,m_3,m_2,m_1)$
to the flavor basis $(\numu,\nut,\nus,\nue)$.
In suggestive notation, 
the $\eps$'s are small angles limited by the short-baseline (SBL) data,
$\thetasun$ and $\thetaatm$
are the angles dominantly responsible for solar and atmospheric oscillations, 
respectively, and $\thetats$ is a possibly large angle 
parametrizing the dominant mixing of the $\nut$ and $\nus$ flavors.
Explicitly, the $\nut$-$\nus$ mixing resulting from $R^T_{23}(\thetats)$
is 
\beq{stmix}
\left(\begin{array}{c}\nup \\ \num \end{array}\right) = 
\left(\begin{array}{cc}
\cos \thetats & -\sin \thetats\\  
\sin \thetats & \cos \thetats \end{array}
\right) 
\left(\begin{array}{c}\nut \\ \nus \end{array} \right).
\eeq

As shown in \cite{Gronau:1985kx},
the three phases may be assigned to the 
double-generation skipping rotations $R_{24}$ and $R_{13}$ and any one
of the single-generation skipping rotations,
in the following manner:
equal and opposite phases are attached to the two non-diagonal elements
of the rotation matrix.  For the single-generation skipping complex rotation, 
we choose $R_{14}$.  
Since the three angles of  $R_{24}$, $R_{13}$, and $R_{14}$ 
are small, this amounts to assigning an arbitrary phase to each of the 
$\sin(\eps_j) \sim \eps_j$'s, with $\eps_j$ to the right of the diagonal
and $\eps_j^*$ to the left.\footnote
{The small-angle expansion of $\cos (\eps_j)$ is
$1-\half |\eps_j|^2$.}
The small-angle part of $U$ is then 
$U_\eps\equiv R_{24}(\epsmm) R_{14}(\epsme) R_{13}(\epsee)$,
which to order $\eps^2$ is
\beq{Ueps}
U_\eps = 
\left(\begin{array}{cccc} 
1-\half (|\epsmm|^2+|\epsme|^2) & - \epsee^* \epsme  & \epsmm & \epsme \\
0 & 1-\half |\epsee|^2 &    0      & \epsee\\
- \epsmm^* & 0 & 1-\half |\epsmm|^2 & -\epsmm^* \epsme \\
-\epsme^* & - \epsee^* & 0 & 1-\half (|\epsee|^2+|\epsme|^2)
\end{array}\right). 
\eeq
We note that with this parameterization the nonzero phases are 
associated exclusively with small angles, making
the smallness of any observable CP-violation in the 2+2 scheme
immediately evident.

When the CP-violating phases are allowed to range from 0 to $\pi$,
all angles may be restricted to the interval $[0,\pi/2]$.
However, when the phases are neglected, the $\theta$'s still range over
$[0,\pi/2]$, but the $\eps$'s now range over $[-\pi/2,\pi/2]$ 
(or over $[0,\pi]$).

The two large-angle mixings on the right in eq.\ \rf{Umatrix}
are given by 
\beq{Unoeps}
U_\pm  = R_{34}(\thetaatm) R_{12}(\thetasun) = 
\left(\begin{array}{cccc} 
\cos \thetaatm & \sin \thetaatm & 0 & 0\\
-\sin \thetaatm & \cos \thetaatm & 0 & 0\\
0 & 0 & \cos \thetasun & \sin \thetasun\\
0 & 0 & -\sin \thetasun & \cos \thetasun\\
\end{array}\right). 
\eeq
This matrix independently mixes each of the two mass doublets in the 2+2
spectrum. $U_\pm$ approximates the full mixing matrix in the 
$(\numu,\nup,\num,\nue)$ basis.
The advantage of locating $U_\eps$ to the left of $U_\pm$ is that
in the full mixing matrix $U$, the angle $\thetaatm$ appears
only in the first two columns, and $\thetasun$ appears only in the last 
two columns.  Consequently, atm/LBL amplitudes do not depend
on $\thetasun$, and solar amplitudes do not depend on $\thetaatm$.
Of course, in any parameterization, the SBL amplitudes 
are insensitive to mixing within either mass doublet, and so 
depend on neither $\thetaatm$ nor on $\thetasun$.

Recent global fits of the 2+2 model 
to solar and atmospheric data  
\cite{Gonzalez-Garcia:2001uy,Maltoni:2001bc,newValle}  
have focused on the seven-parameter set 
$\{\epsmm, \thetats, \thetasun, \thetaatm, \dmlsnd, \dmsun, \dmatm \}$, 
neglecting the two small angles $\epsme, \epsee$,
and the three CP-violating phases.\footnote
{In ref.\ \cite{newValle} the angles $\epsme$ and $\epsee$ are 
turned on, but only in the LSND part of the analysis.
}
The investigation of the dependence of the sterile neutrino sum rule
on the small angles ($\eps$'s) in this paper suggests that neglect 
of these small angles, especially $\epsme$, may not be warranted.

\section{Sum and Product Rules}
\subsection{Zero$^{th}$ order in $\eps$'s}

Let us first discuss the oscillation probabilities in vacuum 
in the limit where
the $\eps$'s are set to zero. This limit results when 
$\langle \nue|\nu_4\rangle =0= \langle \nue|\nu_3\rangle$, and  
$\langle \numu|\nu_2\rangle =0= \langle \numu|\nu_1\rangle$.

In the limit of vanishing $\eps$'s, 
one can use eq.\ \rf{Unoeps} 
to read off immediately that 
LBL/atm oscillations of the $\nue$ are zero and 
$\numu$ oscillates into pure $\nu_+$;
at the solar-scale, $\nue\rightarrow \numu$ oscillations are zero,  
and $\nue$ oscillates into pure $\nu_-$;
and there are no SBL oscillations.

It proves useful to denote the amplitude of the CP-conserving 
oscillation at each $\delta m^2$ scale 
(short-baseline/LSND, long-baseline/atmospheric, and solar) 
by \cite{Barger:1998bn}

\beq{amps}
A_{\rm SBL}(\nu_\alpha\rightarrow\nu_\beta)
=-4\sum_{j=1}^2\sum_{k=3}^4
R^{\alpha j}_{\beta k}\,;\quad
A_{\rm LBL}(\nu_\alpha\rightarrow\nu_\beta)
=-4R^{\alpha 3}_{\beta 4}\,;\quad
A_{\rm sun}(\nu_\alpha\rightarrow\nu_\beta)
=-4R^{\alpha 1}_{\beta 2}\,.
\eeq
The disappearance amplitude is denoted by
\beq{disamp}
A_\Delta (\nu_\beta\rightarrow\nu_{\not\beta}) 
=\sum_{\alpha\neq\beta} A_\Delta (\nu_\beta\rightarrow\nu_\alpha)
= 4\sum_\Delta |U_{\beta j}|^2\,|U_{\beta k}|^2\,;
\eeq
here, $\Delta$ denotes the SBL, LBL/atm, or solar scale, and 
$\sum_\Delta$ denotes the sum over the appropriate 
mass states indicated explicitly in eq.\ \rf{amps}.
Similarly, the CP-conserving contribution to the probability 
at each $\delta m^2$ scale is denoted by
\beq{probs}
P_\Delta (\nu_\alpha\rightarrow\nu_{\beta}) \equiv
A_\Delta (\nu_\alpha\rightarrow\nu_{\beta}) 
\sin^2\left(\frac{\delta m^2_\Delta\,L}{4E}\right)\,.
\eeq
Some care is required in applying these contributions to 
oscillation data,
since observables at any given scale will in general receive 
contributions from that scale and all shorter scales.
Such is not the case with the block-diagonal mixing 
of eq.\ \rf{Unoeps}.

Explicitly, the nonzero oscillation amplitudes for $\numu$ due to 
atmospheric-scale oscillations are 
\ba{noepsatm}
A_{\rm atm}(\numu \rightarrow \nut) &=&  \sin^2 (2\thetaatm)\, \cos^2 
\thetats
\label{noepsatm} \\
A_{\rm atm}(\numu \rightarrow \nus) &=& \sin^2 (2\thetaatm)\, \sin^2 
\thetats\\
A_{\rm atm}(\numu \rightarrow \nunotmu) &=&  \sin^2 (2\thetaatm)
\ea
while for $\nue$ due to solar-scale oscillations they are
\ba{noepssun}
A_{\rm sun}(\nue \rightarrow \nut) &=& \sin^2 (2\thetasun)\, \sin^2 
\thetats\\
A_{\rm sun}(\nue \rightarrow \nus) &=& \sin^2 (2\thetasun)\, \cos^2 
\thetats \label{noepssun}\\
A_{\rm sun}(\nue \rightarrow \nunote) &=& \sin^2 (2\thetasun)
\ea

We note that matter effects cannot alter the texture of the 
off-diagonal elements of the  
mass-squared matrix in the flavor basis, because the matter 
potential is diagonal in this basis.
Consequently, the block-diagonal structure of the diagonalizing matrix
in eq.\ \rf{Unoeps} is maintained in the presence of matter.
This further implies that eqs.\ \rf{noepsatm} to 
(\ref{noepssun})
are unchanged in form by matter, 
although the angles (and mass-eigenvalues)
assume different values.

In solar matter, there is an essential nuance enriching the 
neutrino physics:
The evolution of the neutrinos from the sun's center
to the sun's surface is dominated by matter effects.
The solar matter-density is not constant, and so the 
matter-dependent mixing-angles change continuously as
the neutrinos transit the sun, possibly effecting 
non-adiabatic transitions among mass-eigenstates. 
However, for large $\theta_{\rm sun}$, which holds for the LMA solution 
which we consider in this work (and for the QVO and VO solutions),
the neutrino propagates adiabatically from the sun's center to 
the surface. 
This means that the neutrino emerges into vacuum with the combination 
of mass eigenstates unchanged from that determined at the sun's center.
Moreover, solar neutrinos with energies above the solar resonance 
associated with $\dmsun$ 
and below the atmospheric resonance 
associated with $\dmatm$ will evolve 
adiabatically to emerge from the sun in a nearly pure 
$\nu_2$ mass-eigenstate.
The solar and atmospheric 
resonant energies in the 2+2 model in the 
zero $\eps$'s approximation are 
\cite{Dooling00} 
\beq{Eressun}
E_{\odot}^R \approx 
\frac{\dmsun\cos (2\thetasun)}{2(A_{CC}-\cos^2 (\thetats) A_{NC})}\,,
\eeq
and 
\beq{Eresatm}
E_{\rm atm}^R \approx 
\frac{\dmatm\cos (2\thetaatm)}{2\sin^2 (\thetats) A_{NC}}\,,
\eeq
respectively.
The vacuum angles and mass-squared differences are defined previously, 
and $A_{CC}=\sqrt{2} G_F N_e$ and $A_{NC}\approx\half A_{CC}$ are 
the effective contributions
of the matter potential to the Hamiltonian in the flavor basis,
written in terms of the electron number density.
Putting in the values $N_e (0)=6.0\times 10^{25} {\rm cm}^{-3}$ 
for the electron density at the center of the sun,
$\dmsun=4\times 10^{-5}{\rm eV}^2$, and 
$\dmatm=3\times 10^{-3}{\rm eV}^2$,
one finds adiabatic evolution to near-pure $\nu_2$ for 
large-angle solar mixing and energies in the range between 
$E_{\odot}^R=4\cos (2\thetasun) (1+\sin^2\thetats)^{-1}$~MeV 
and $E_{\rm atm}^R=300 \cos(2\thetaatm) (\sin^2\thetats)^{-1}$~MeV.
The solar neutrinos from the $^8 B$ decay chain,
measured in the SuperK and SNO experiments,
lie within this energy range.

The neutrinos arriving at earth from the sun 
are incoherent mass eigenstates.\footnote
{Even if the neutrinos exiting the sun bear coherent phase 
relationships, they decohere in transit to Earth if 
$\dmsun \gsim 10^{-10}{\rm eV}^2$,
due to the many oscillation lengths contained in the 
one A.U. path length.
}
Thus, the density operator describing 
the decoherent ensemble of neutrinos arriving at earth from the sun
is diagonal in the mass basis.
In the adiabatic approximation,
it is simply 
\beq{rhoMass}
{\hat \rho}^{\rm S}_{\rm mass} =  \sum_j |U^S_{ej}|^2 \;|\nu_j><\nu_j|\,.
\eeq
Here, $U^S$ is the mixing matrix at the center of the sun where 
the solar neutrinos originate.  
Let us label the solar neutrino ``$\nusun$'' to signify its 
$\nue\rightarrow\nu_{SUN}$ solar history.
The probability to measure neutrino flavor $\beta$ at earth
is then given by
\beq{Psun}
P_S (\nusun\rightarrow\nu_\beta) = 
<\nu_\beta\,|{\hat \rho}^S|\,\nu_\beta>
          = \sum_j |U^S_{ej}|^2 \;|U^V_{\beta j}|^2 \,,
\eeq
after applying eqs.\ \rf{U} and \rf{Udag}.
The $\nusun\equiv \nu_2$ approximation\footnote
{In our numerical work we will go 
beyond the $\nusun\equiv \nu_2$ approximation.
Nevertheless, $\nusun\equiv\nu_2$ is a good approximation 
offering simple results.  Numerically, we find that 
$|U^S_{e2}|^2 = 90\%$, 98\%, and 99\% for $E_\nu=5$, 10, and 15~MeV,
respectively, for zero $\eps$'s, and with little change for 
nonzero $\eps$'s.
}
consists of setting 
$|U^S_{ej}|^2$ equal to $\delta_{j2}$, yielding 
\beq{nu2approx}
\quad\quad\quad\quad 
 P_{\nu_\odot \rightarrow \nu_\beta} = |U^V_{\beta 2}|^2
\quad\quad\quad\quad 
 [\nusun\equiv\nu_2,\;{\rm adiabatic}\,]
\eeq
In the absence of $\eps$'s, we have from eqs.\ \rf{stmix}, 
\rf{Unoeps}, and \rf{nu2approx}, that 
\beq{nu2}
\nu_2=\cos\theta_{\rm sun} 
(\sin\theta_{\tau s}\nut +\cos\theta_{\tau s}\nus)
-\sin\theta_{\rm sun}\,\nu_e\,.
\eeq
>From this equation, we may read off the values of 
$U^V_{\beta 2}$ to obtain the oscillation 
amplitudes for solar neutrinos arriving at the earth
in the $\nusun \equiv \nu_2$ approximation:
\ba{noeps_sun}
A_{\rm sun}(\nusun \rightarrow \nut) &=& \cos^2 (\thetasun)\, \sin^2 
\thetats\\
A_{\rm sun}(\nusun \rightarrow \nus) &=& \cos^2 (\thetasun)\, \cos^2 
\thetats\\
A_{\rm sun}(\nusun \rightarrow \nunote) &=& \cos^2 (\thetasun)
\ea
In these formulas, the angles are truly vacuum angles.

Thus, unitarity in the context of the 2+2 mass spectrum has led one to a
sum rule \cite{Peres:2001ic} and a product rule at zero$^{th}$ order in $\eps$:

\beq{SR1}
\left[\frac{P(\nusun \rightarrow \nus)}{P(\nusun \rightarrow \nunote)}
\right]_{\rm sun}
+\;
\left[\frac{P(\numu \rightarrow \nus)}{P(\numu \rightarrow \nunotmu)}
\right]_{\rm atm}
= (\cos^2 \thetats)_V +(\sin^2 \thetats)_E
\rightarrow 1
\label{SR1}
\eeq
and

\beq{SR2}
\left[\frac{P(\nusun \rightarrow \nus)}{P(\nusun \rightarrow \nut)}
\right]_{\rm sun}
\times \;
\left[\frac{P(\numu \rightarrow \nus)}{P(\numu \rightarrow \nut)}
\right]_{\rm atm}
=(\cot^2\thetats)_V \times (\tan^2\thetats)_E
\rightarrow 1
\label{SR2}
\eeq
where the subscripts $V$ and $E$ signify ``vacuum'' and ``earth-matter'',
and the arrows hold in the limit of negligible earth-matter effects.
In the figures to follow and the discussion thereof 
we will refer to the ratio terms in eqs\ (\ref{SR1}) and (\ref{SR2})
as $\Rsun$ and $\Ratm$, respectively. 

We emphasize that the sum and product rules are not
required by any underlying symmetry or principle.
Rather, they are accidents of the block-diagonal structure of 
eq.\ \rf{Unoeps}, which results when the small angles are neglected.
The values on the right-hand sides of these two rules, 
eqs.\ \rf{SR1} and \rf{SR2}, will differ from unity when 
the small mixing angles are included, even if matter-effects 
are absent.  Furthermore, although their values 
are intimately related at zero$\th$ order in the $\eps$'s,
they are not so simply related at higher order in $\eps$'s.

The purpose of this paper is to evaluate these sum rules, 
including the small angles neglected in previous 
work, and including possible earth-matter effects.
In the next section, we calculate oscillation probabilities 
to order $\eps^2$,
to find the order $\eps^2$ corrections to the sum 
\rf{SR1} and product \rf{SR2} rules.
In section \ref{sec:results} we numerically evaluate the sum and 
product rules without approximation.
Readers interested only in the answers may jump over 
the remaining formalism to find the results presented in 
section \ref{sec:results}.

\subsection{Order $\eps$ and $\eps^2$ corrections}
To order $\eps^2$, the full mixing matrix in the physical flavor-basis
$(\numu,\nut,\nus,\nue)$ is the product 
\beq{Ufull}
U=
\left(
\begin{array}{cccc}
1 & 0& 0& 0\\
0 & \cos \thetats & \sin \thetats & 0\\  
0 & -\sin \thetats & \cos \thetats & 0\\
0 & 0 & 0 & 1 
\end{array}
\right) 
\times U_\eps
\times U_\pm  
\eeq
where $U_\eps$ and $U_\pm$ are given in 
eqs.\ \rf{Ueps} and \rf{Unoeps}.
This mixing matrix determines all of the oscillation 
amplitudes to ${\cal O}(\eps^2)$, and so determines the 
${\cal O}(\eps)$ and ${\cal O}(\eps$$^2)$
modifications to the sum rules.
Here we give the individual flavor-changing amplitudes 
for SBL, solar, and atmospheric neutrinos to ${\cal O}(\eps^2)$.
We begin with the SBL amplitudes, and summarize the 
existing experimental constraints on these amplitudes.

In our notation, the SBL oscillation amplitudes to order $|\eps|^2$ are 
\ba{SBLamps}
A_{\rm SBL}(\nue \rightarrow \nu_{\not e}) &=& 
4\,[|\epsme|^2 +|\epsee|^2]
\label{SBLamp1}\\
A_{\rm SBL}(\numu \rightarrow \nu_{\not\mu}) &=&
4\,[|\epsme|^2 +|\epsmm|^2]
\label{SBLamp2}\\
A_{\rm SBL}(\numu \rightarrow \nue) &=& 4\,|\epsme|^2
\label{SBLamp3}\\
A_{\rm SBL}(\numu \rightarrow \nut) &=& 
4 |\epsmm|^2 \sin^2 \theta_{\tau s}
\label{SBLamp4}\\
A_{\rm SBL}(\numu \rightarrow \nus) &=&
4 |\epsmm|^2 \cos^2 \theta_{\tau s}  
\label{SBLamp5} \\
A_{\rm SBL}(\nue \rightarrow \nut) &=& 
4|\epsee|^2 \cos^2 \theta_{\tau s}
\label{SBLamp6}\\
A_{\rm SBL}(\nue \rightarrow \nus) &=& 
4|\epsee|^2 \sin^2 \theta_{\tau s}.
\label{SBLamp7}
\ea
These amplitudes, 
and so the values of the small angles $\epsee$, $\epsme$ and $\epsmm$,
are bounded from above  
by the experimental limits on $\numu$ and $\nue$ disappearance in vacuum,
and by atmospheric neutrino oscillation data.
The positive result of LSND bounds $\epsme$ from below,
but this constraint is not significant for present purposes.
For the allowed LSND region 
$\delta m_{\rm SBL}^2\sim 1.0\,{\rm to}\,0.2~{\rm eV}^2$,
the BUGEY disappearance experiment \cite{Declais:1994su} 
provides the 90 \% C.L. bound\footnote{
The KARMEN experiment provides a tighter bound than the BUGEY experiment
for $\dmlsnd > 0.2 {\rm eV}^2$, reaching 
$\frac{1}{4} A_{\rm SBL}(\numu \rightarrow \nu_e)\sim 0.7\times 10^{-3}$
at $\dmlsnd\sim 1 {\rm eV}^2$.
However, atmospheric experiments average over the SBL contribution,
and so are not sensitive to the value of $\dmlsnd$.
Accordingly, the appropriate bound to use is the more liberal 
BUGEY bound, inferred at $\dmlsnd\sim 0.2 {\rm eV}^2$.
}
$\frac{1}{4} A_{\rm SBL}(\nue \rightarrow \nu_{\not e})
= |\epsee|^2+|\epsme|^2 \le 0.01$.
The CDHS \cite{Dydak:1983zq}
$\bar{\nu}_\mu$ ($\nu_\mu$) disappearance experiment
bounds the amplitude 
$\frac{1}{4} A_{\rm SBL}(\numu \rightarrow \nu_{\not \mu}) =
|\epsmm|^2+|\epsme|^2 \le 0.2$
for $\delta m^2_{\rm SBL} \gsim 0.3\,{\rm eV}^2$.
In fact, a more stringent bound on 
$A_{\rm SBL}(\numu \rightarrow \nu_{\not \mu})$
results from atmospheric neutrino data
(a nonzero value for this amplitude is incompatible 
with maximal $\numu$ mixing at the $\dmatm$ scale).
A fit to atmospheric data \cite{Gonzalez-Garcia:2001uy} results in
$A_{\rm SBL}(\numu \rightarrow \nu_{\not \mu})<0.48\,(0.64)$ with
90\% (99\%) C.L., which translates into 
$|\epsmm|^2+|\epsme|^2 < 0.12\,(0.16)$.
We note that the CHOOZ limit on $\nue$-disappearance 
at the atmosphere scale is of order $(\epsme,\,\epsee)^4 \lsim 10^{-4}$ 
in the context of the 2+2 model,
and so is not of interest.

To summarize these constraints, the 90\% C.L. 
bounds which we will use in our numerical
evaluation of the sum rules are:
(i) $|\epsee|^2+|\epsme|^2 \le 0.01$, and 
(ii) $|\epsmm|^2 < 0.12$.
Note that these bounds allow $\epsee$ and $\epsme$
to be as large as 0.10,
and allow $\epsmm$ to be as large as 0.35.


Before ending this discussion of the SBL amplitudes, 
we note in passing a new product rule for 
SBL amplitudes.  Emerging from  eqs.\ (\ref{SBLamp4})-(\ref{SBLamp7}) 
is
\beq{SR3}
\left[\frac{P(\nue \rightarrow \nus)}{P(\nue \rightarrow \nut)}
\right]_{\rm SBL} \times \;
\left[\frac{P(\numu \rightarrow \nus)}{P(\numu \rightarrow \nut)}
\right]_{\rm SBL}=1\,,
\eeq
exact to order $\eps^2$, at least.

Next we turn to the amplitudes for solar neutrino oscillations.
We employ again the adiabatic approximation valid for large 
solar-mixing solutions, and simplify the discussion here with 
the $\nusun\equiv\nu_2$ approximation.
To ${\cal O}(\eps^2)$, the expansion of $\nu_2$ in flavor states is 
\beq{adiabatic}
|\nu_2> = \left\{
\begin{array}{lr}
 & (\epsmm\cos\thetasun -\epsme\sin\thetasun)\,|\numu> \\
+& (\sin\thetats[\cos\thetasun (1-\half|\epsmm|^2) +
    \sin\thetasun \epsmm^* \epsme] -\cos\thetats\sin\thetasun\epsee)\,
    |\nut> \\
+& (\cos\thetats[\cos\thetasun (1-\half|\epsmm|^2) +
    \sin\thetasun \epsmm^* \epsme] +\sin\thetats\sin\thetasun\epsee)\,
    |\nus> \\
-& \sin\thetasun (1-\half|\epsee|^2-\half|\epsme|^2)\,|\nue>.
\end{array}
\right.
\eeq
For solar neutrinos arriving at the ``day'' hemisphere of the earth,
all angles assume vacuum values.
This result generalizes eqn.\ \rf{nu2} to nonzero $\eps$'s.
>From this, one can easily calculate the first term in the sum rule
\rf{SR1},
\beq{SR1sun}
\left[
\frac{P(\nusun\rightarrow\nus)}{P(\nusun\rightarrow\nunote)}
\right]_{\rm sun} 
\simeq \;
\frac{|<\nus|\nu_2>|^2}{1-|<\nue|\nu_2>|^2}\,,
\eeq
as a function of the large and small angles.
Similarly, the first factor in the product rule \rf{SR2}, 
\beq{SR2sun}
\left[\frac{P(\nusun \rightarrow \nus)}{P(\nusun \rightarrow \nut)}
\right]_{\rm sun} 
\simeq \;
\frac{|<\nus|\nu_2>|^2}{|<\nut|\nu_2>|^2}\,,
\eeq
can be readily calculated.

Finally, we turn to the oscillation amplitudes for the atmospheric 
neutrinos.
The second terms in the sum and product rules, namely,
\beq{SR1atm}
\left[\frac{P(\numu\rightarrow\nus)}{P(\numu\rightarrow\nunotmu)}
\right]_{\rm atm} = \;
\frac{P_{\rm LBL}(\numu\rightarrow\nus)+P_{\rm SBL}(\numu\rightarrow\nus)}
      {P_{\rm LBL}(\numu\rightarrow\nunotmu)+P_{\rm SBL}(\numu\rightarrow\nunotmu)}\,,
\eeq
and
\beq{SR2atm}
\left[\frac{P(\numu \rightarrow \nus)}{P(\numu \rightarrow \nut)}
\right]_{\rm atm} = \;
\frac{P_{\rm LBL}(\numu\rightarrow\nus)+P_{\rm SBL}(\numu\rightarrow\nus)}
      {P_{\rm LBL}(\numu\rightarrow\nut)+P_{\rm SBL}(\numu\rightarrow\nut)}\,,
\eeq
respectively, are given by inputting the appropriate amplitudes. 
In the atmospheric data, the measurement process averages over
oscillation scales small relative to the Earth's radius.  
Hence it is correct and 
necessary to include contributions from the LBL 
length-scale and smaller, which here includes the oscillation-averaged
short baseline amplitudes.
One notes that the SBL amplitudes, given in eqs.\ 
(\ref{SBLamp1})-(\ref{SBLamp7}),
contribute to the sum rules 
at order $\eps^2$ but not order $\eps$.
The long baseline amplitudes to order $\eps^2$ are: \\
\ba{LBLeps2}
A_{\rm LBL}(\numu \rightarrow \nus) &=&\sin^2 2 \thetaatm\,[
\sin^2 \theta_{\tau s} ( 1 - |\epsme|^2 -|\epsee|^2)
- |\epsmm|^2] \label{numslbl} \nn \\
&& -  \sin 4 \thetaatm [\frac{1}{2} \sin 2 \theta_{\tau s} Re(\epsmm)
+ Re(\epsmm \epsme^*)
\sin^2 \theta_{\tau s}]
\\
A_{\rm LBL}(\numu \rightarrow \nut) &=&\sin^2 2 \thetaatm\,[
\cos^2 \theta_{\tau s}( 1 - |\epsme|^2 -|\epsee|^2)
- |\epsmm|^2] \nn \\
&& +  \sin 4 \thetaatm [\frac{1}{2} \sin 2 \theta_{\tau s} Re(\epsmm)
- Re(\epsmm \epsme^*)
\cos^2 \theta_{\tau s}]
\\
A_{\rm LBL}(\numu \rightarrow \nue) &=&-\sin 2 \thetaatm\,[
(|\epsme|^2 - |\epsee|^2) \sin 2 \thetaatm + 2 Re(\epsme^*
\epsee) \cos  2 \thetaatm] \\
A_{\rm LBL}(\numu \rightarrow \nu_{\not\mu}) &=&
\sin^2 2 \thetaatm\,[ 1 - 2|\epsme|^2 - 2|\epsmm|^2]
- 2 \sin 4 \thetaatm Re(\epsmm \epsme^*)
\ea
With eqs.\ \rf{SR1sun}-\rf{SR2atm} as our guide, 
it is not difficult to write out the explicit analytic expressions
for the sum and product rules.  It is also not
especially illuminating to do so.
One finding is that the order $\eps$ and $\eps^2$ corrections are different
for the sum and product rules.  Thus these two rules,
containing the same information in zero$^{\rm th}$ order,
contain different information when the small angles are included.
We remind the reader that atmospheric oscillations occur in the earth,
and so the mixing angles that enter eqs.\ \rf{SR1atm} and \rf{SR2atm}
are matter rather than vacuum angles.

\section{MATTER EFFECTS}
As has already been discussed,
the effect of solar matter on the emerging solar neutrinos 
is very significant.  
Turning to earth-matter, 
we find in our numerical work that its effects are small at low neutrino
energies $\sim 1$~GeV, but large at high energies.
Thus at low energy, the neutrino plots we show are little changed
for the atmospheric antineutrino flux.
Among the energy bins we consider,
the largest difference for neutrino vs.\ antineutrino occurs in the 
1.5 to 30~GeV bin.
At still higher energies, 
we find the main effect of matter is a suppression
of oscillation, common to both neutrinos and antineutrinos,
and an interesting resonance at the $\dmsq$ scale, in the 
$\numu-\nu_+$ channel \cite{Dooling00}.  
This resonance is independent of the 
sign of the matter potential for large $E\,\delta A_{NC}/\dmatm$,
and so it too is common to both neutrinos and antineutrinos.
Thus, we find that overall the sum rule results we show for 
neutrinos are little altered for antineutrinos.

The relevant part of the neutrino Hamiltonian 
in the flavor basis is
\beq{Hflavor}
H^\kappa_F=U_V\,\frac{M^2_V}{2E}\,U^\dag_V + A_\kappa\,.
\eeq
The matter-induced potentials are  
$A=\half\sqrt{2}G_F\,{\rm diag}(-N_n,-N_n,\,0,\,2N_e-N_n)$ 
in the $(\numu, \nut, \nus, \nue)$ flavor-basis;
$N_e$ and $N_n$ are the electron and neutron densities.
For antineutrinos, the sign of $A$ is reversed. 
Since an overall phase is not measurable, it is useful to 
translate $A$ by $+N_n {\bf 1}$ and use the relative potentials 
$A=\sqrt{2}G_F\,{\rm diag}(0,\,0,\,\half N_n,\,N_e)$. 
With $N_e \approx N_n$, which holds in the earth and in the sun,
one has $A\equiv (0,0,A_{NC}, A_{CC})$, with 
$A_{CC}= \sqrt{2}G_F N_e \approx 2A_{NC}$,
as presented earlier.
Numerically, $A_{CC}=0.80\times 10^{-13}(N_e/N_A{\rm cm}^{-3})$~eV.
Electron densities in the earth's mantle and core are
$(N_e)_M \approx 1.6\,N_A/{\rm cm}^3$ and 
$(N_e)_C \approx 6\,N_A/{\rm cm}^3$.

Diagonalization of \rf{Hflavor} to 
\beq{Hdiag}
H^\kappa_{(\rm diag)}=U^\dag_\kappa\,H^\kappa_F\,U_\kappa
\eeq
is done numerically in our work.
It yields the mixing-matrix in matter ($U_\kappa$) and the 
mass-squared eigenvalues in matter 
(the diagonal elements of $H^\kappa_{(\rm diag)}$).
The matter contribution to the Hamiltonian can significantly 
alter the masses and mixing angles, compared to their vacuum values.

Insight into the multi-resonance possibilities 
among four neutrino flavors may be found by using the approach of 
\cite{Dooling00}. The idea is to find a neutrino basis 
such that the effective Hamiltonian matrix, 
including matter potentials,
nearly decouples into a resonant $2\times 2$ sub-block embedded in the rest.
Decoupling may occur when $E\delta A$ is matched to just one of
a hierarchy of $\dmsq$ values.
A two-neutrino analysis may then be applied to 
the decoupled $2\times 2$ sub-block.

Consider the $2\times 2$ effective Hamiltonian (hermitian) matrix 
\beq{twobytwo}
H=
\left(\begin{array}{cc} 
h_{22}   & h_{21} \\
h_{21}^* & h_{11} \\
\end{array}\right)\,.
\eeq
The rotation angle which diagonalizes $H$ is given by
\beq{thetam}
\tan(2\thetam)=\frac{2|h_{21}|}{h_{22}-h_{11}}\,.
\eeq
The oscillation amplitude is $\sin^2(2\thetam)$.
The eigenvalue difference is 
\beq{hdiff}
\delta h = \sqrt{(h_{22}-h_{11})^2 +4|h_{21}|^2}=2E\delta m^2_M\,,
\eeq
where the second equality defines the effective mass-squared difference 
$\delta m^2_M$ in matter.

It often happens that the denominator $h_{22}-h_{11}$ 
varies from positive through zero 
to a large negative value as energy or matter density are varied.
When this is so, $\thetam$ varies from its vacuum value $\theta_V$, 
through resonance and maximal mixing at $\pi/4$, 
ultimately to nearly $\pi/2$.
The resonance with $\thetam=\pi/4$ and maximal oscillation amplitude 
$\sin^2(2\thetam)=1$ occurs when $h_{22}=h_{11}$.
The resonance width, defined by the half-maximum amplitude 
$\sin^2 (2\thetam)=\half$, i.e. $\tan^2 (2\thetam)=1$,
is given by $|h_{22}-h_{11}|=2|h_{21}|$.

For comparison purposes, it is useful to recall the 
exact two-flavor results.
The oscillation amplitudes in matter and vacuum are 
related by
\beq{twoamp}
\sin^2 (2\thetam)=
\frac{\sin^2 (2\thetav)}{\sin^2 (2\thetav)+\xi}\,,
\eeq
where $\xi$ is defined by
\beq{xi}
\xi\equiv \left(\cos(2\thetav)-\frac{2E\delta A}{\dmsq}\right)^2
= \cos^2(2\thetav)\,\left(1-\frac{E}{E_R}\right)^2. \,,
\eeq
Here, $\delta A= A_1 - A_2$ is the difference in matter potentials 
ordered {\sl opposite} to the difference in vacuum mass-squares,
$\dmsq$, and the resonant energy in the two-flavor system is 
\beq{Eres}
E_R=\frac{\dmsq\cos (2\thetav)}{2\delta A}
	=10\cos (2\thetav)  
	\left(\frac{\dmsq}{2\times 10^{-3}{\rm eV}}^2\right)
	\left(\frac{10^{-13}\,{\rm eV}}{\delta A}\right)\,{\rm GeV}\,.
\eeq
Mass-squared differences in vacuum and matter are related by 
\beq{twomass}
\dmsq_M = \dmsq\,\sqrt{\sin^2 (2\thetav) +\xi}\,.
\eeq
The oscillation length is changed from its vacuum value 
$\lambda_V=2.48\,(E/{\rm GeV})({\rm eV}^2/\delta m^2)$~km to
\beq{lambdam}
\lambda_M=\frac{\dmsq}{\dmsq_M}\lambda_V 
	  = \frac{\lambda_V}{\sqrt{\sin^2 (2\thetav) +\xi}}\,.
\eeq
Note that in the two-flavor analysis,
a near maximal vacuum angle forces the resonant energy toward zero
(eq.\ \rf{Eres},
and further, necessarily implies suppression 
of the amplitude (eq.\ \rf{twoamp}) and contraction of the oscillation 
length (eq.\ \rf{lambdam}).
The atmospheric vacuum mixing-angle is nearly maximal.

Note also the trend as the energy $E$ is increased:
$\xi$ varies from $\cos^2 (2\thetav)$ at low energy,
to zero at resonance, to a value increasing as $E^2$ above 
resonance.
At the resonant energy, the oscillation amplitude
$\sin^2 2\thetam$ attains its maximum, unity,
the mass-squared difference attains its minimum,
$(\delta m_M^2)_R= \delta m^2\, \sin (2\theta_V)$,
and the oscillation length is
$\lambda_R=\lambda_V/ \sin (2\theta_V)$.
For small vacuum angles, these changes are large.
With large mixing angles, the changes are not dramatic;
in addition,
the resonant energy is very small and the resonance is broad.
The sign of $\delta m^2/\delta A$ determines whether the resonances 
occur in the neutrino or antineutrino channel.
For $\theta_V \le 45^{\circ}$ (the ``light side''),
the neutrino (anti-neutrino) channel is resonant for 
$\delta A$ as defined and $\delta m^2$
of same (opposite) sign.
Equivalently,
resonant oscillations occur in the neutrino channel when 
$E_R$ as defined in \rf{Eres} is positive,
and in the anti-neutrino channel when $E_R$ as defined is negative.  
As can be seen from eq.\ \rf{twoamp}, large $E$ suppression 
of oscillations is the same for neutrino and antineutrino.

For neutrino energies well above the resonant value,
the oscillation amplitude is suppressed as $E^{-2}$,
and the oscillation length is contracted toward its
energy-independent asymptotic value
$\lambda_M =2\pi/|\delta A|$.  Numerically, this 
asymptotic oscillation length is
$\lambda_M = 1.3\,(10^{-13}{\rm eV}/|\delta A|)\times 10^4$~km,
which is of order of the Earth's diameter $D_\oplus$ 
for the values of $A_{NC}$ in earth matter.
Before approaching its asymptotic value, 
the oscillation length reaches a maximum value of 
$\lambda_{\rm max}=2\pi/|\delta A|\sin2\theta_V$.
Thus, with just two flavors the main effect of matter at large energy 
is to shorten 
oscillations relative to their vacuum values, but especially to  
suppress oscillations as $\thetam$ is pushed toward its  
asymptotic value of $\pi/2$.

In contrast, with four flavors, the six rotations and additional 
matter potential greatly enrich the matter-dependent oscillations.
Diagonalizing the effective $2\times 2$ sub-blocks gives results 
qualitatively different from diagonalizing an exact $2\times 2$
matrix in a true flavor basis.  We find, for example, that in the 
energy range $\dmatm/A_{NC}\ll E\ll \dmlsnd/A_{NC}$ the atmospheric 
oscillation probability for both $\numu$ and $\numubar$ may be 
significantly enhanced by matter,
when the small vacuum angles satisfy certain relations.
This surprising result is contrary to the suppression one expects 
from studying a pure two-flavor model. 
We reserve the detailed analysis of particular oscillation-channels
for a future publication \cite{future}, since the focus of 
this (already long) paper is the sum rule, which involves 
ratios of oscillation channels.

\section{RESULTS}
\label{sec:results}
\subsection{Zero$\th$ Order}
\label{sub:zeroorder}
\begin{figure}
\begin{center}
\epsfxsize=110mm
\hspace*{1cm}
\epsfbox{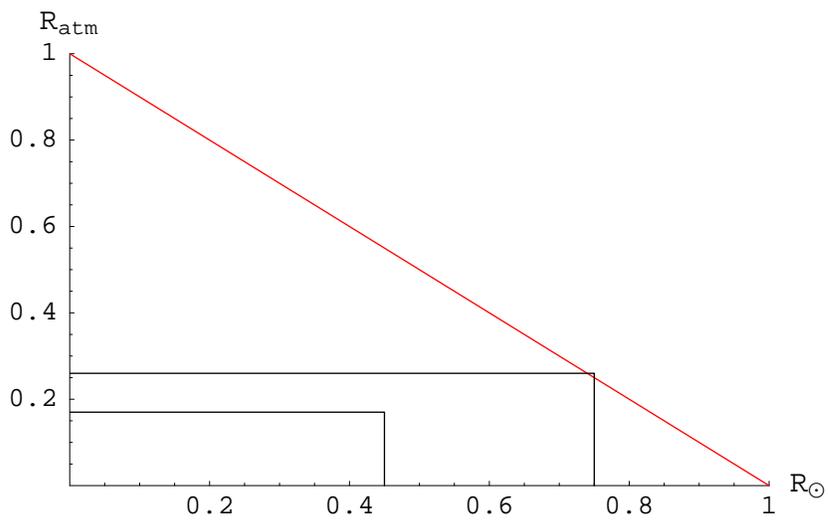}
\end{center}
\caption{The zero$\th$ order sum rule compared to the
90\% and 99\% exclusion boxes obtained from fits to data 
with small angles set to zero.
The vertical axis is $R_{\rm atm}$ and the horizontal axis is $R_{\rm sun}$.
\label{fig:zero}}
\end{figure}

We begin our presentation of results 
with a plot (Fig.\ \ref{fig:zero}) 
of the zero$\th$ order sum rule, and the 
exclusion boxes that result when all small angles are 
set to zero \cite{Gonzalez-Garcia:2001uy}.
The axes labels are 
\beq{Ratm}
R_{\rm atm}\equiv \left[
\frac{P(\numu\rightarrow\nus)}{P(\numu\rightarrow\nunotmu)}
\right]_{\rm atm}
\eeq
and
\beq{Rsun}
R_{\rm sun}\equiv \left[
\frac{P(\nu_\odot\rightarrow\nus)}{P(\nu_\odot\rightarrow\nunote)}
\right]_{\rm sun}\,.
\eeq
The 90\% exclusions are 0.17 for the sterile contribution 
to the atmospheric data,
and 0.45 for the sterile contribution to the solar data.
The 99\% exclusions are 0.26 and 0.75, respectively.
A similar value of $R_{\rm sun}<0.7$ at 3 $\sigma$ is given in
\cite{snosterile}).
We have checked that the effect of earth matter on the sum and 
product rules is 
negligible (less than 1\%) when the small angles are zero.
As a consequence, the zero$\th$ order plot of Fig.\ \ref{fig:zero} 
stands as a prediction for  
oscillation data over the entire relevant range of atmospheric neutrino 
energies, 0.5~GeV to 500~GeV, when small angles are neglected.
In the context of eqs.\ \rf{SR1} and \rf{SR2},
this means that the change in $\thetats$ 
due to earth-matter is negligible in the whole 
$[0.5,\,500]$~GeV energy region when $\eps$'s are set to zero.

As one sees in Fig.\ \ref{fig:zero},
the overt sterile of the 2+2 sum rule conflicts with the 
sterile exclusion bounds coming from the solar and atmospheric data
when the three small angles of the 2+2 model are neglected.
The 99\% exclusion box just touches the sum-rule line at a single point,
where $\tan^2\thetats=R_{\rm atm}/R_{\rm sun}=0.26/0.75$,
or $\thetats=30^\circ$.
The question arises whether inclusion of these small angles
can sufficiently scatter the sum rule and weaken the exclusion boxes
to make the 2+2 model viable.
Re-calculating the exclusion boxes with nonzero small angles
is a formidable task beyond the scope of this paper.
We note, however, that a global fit with the largest of the small
angles ($\epsmm$ in our notation) included has been performed. 
The consequence is an expanded exclusion box
for the value of $\cos^2 \theta_{\tau s} \cos^2 \epsmm$
derived from atmospheric neutrinos. 
However this latter quantity is not directly related to
$R_{\rm atm}$ for $\epsmm \neq 0$,
and so we continue to show just the restricted fit ($\eps$'s = 0) 
in the figures to follow.
Nevertheless, 
the trend is clear and expected: 
adding the small angles allows 
a larger sterile neutrino component in the global data.
 
\subsection{All Orders}
\label{sub:allorders}
We next show that the sum rule is significantly broadened 
when small angles are included.

We display our numerical results,
which include the effects of the nonzero small-angles and matter,
in scatter plots in the plane labeled by the two terms of the
sum rule or product rule.  
Each scatter point is determined by a random choice of the 
three small angles within their allowed regions,
and a random choice of $\thetats$ 
within the interval $[0,\frac{\pi}{2}]$.
In the numerical work, 
we use the full rotation matrices for the small angles,
rather than the small-$\eps$ expansion presented earlier.
We allow each $\eps$ to range over
$-\eps_{\rm XB}\leq\eps\leq \eps_{\rm XB}$,
where the set of $\eps_{\rm XB}$ are the experimental bounds derived
earlier in this paper (either 0.10 or 0.35).  
We remind the reader that we must include negative values of 
$\eps$ to (partially) atone for our neglect of the CP-violating 
phases independently associated with each $\eps$.  
We do not take on the extra 
burden of calculating with complex $\eps$'s,
as this would double the number of small-angle parameters, 
from three to six.
For the atmospheric data sample, 
we restrict ourselves to the angular bin that is most up-going
($-1 \le \cos\theta_z \le -0.8$).
The most up-going bin is chosen because 
events in this bin have the longest baseline 
and so have enhanced oscillation probabilities, and  
the best experimental test of the sterile/active neutrino ratio 
in atmospheric data involves the matter-effect.

We simulate the low-energy atmospheric data set by averaging the 
incident neutrino energy over a specific energy range,
and by averaging the zenith angle over 
$-1 \le \cos\theta_z \le -0.8$, for each scatter point.
Four specified energy ranges are chosen.  They are 
0.5 to 1.5~GeV, to simulate the contained events,
1.5 to 30~GeV, to simulate the partially contained events,
30 to 500~GeV, to simulate the full range of through-going events,
and 50 to 150~GeV, to display the ``typical'' energy of 
through-going events.
Each point displayed in the scatter plot is constructed
as the mean 
value of 500 events, each a product of 50 randomly chosen energy
values times 10 randomly chosen zenith-angle values.
To judge the stability of this averaging, we repeated some trials
with 500 energy values times 100 zenith-angle values.
The results differed by less than 1\%. 

For the solar data, we average over the energy range 5 to 15 MeV,
by choosing 50 random points within this range.

The following numerical values of
the solar LMA
\footnote{After publication of the SNO NC data the LMA solution is 
selected at 99 \% C.L. \cite{snolma}}
and atmospheric best fits \cite{Gonzalez-Garcia:2001uy}
are taken as fixed:
\ba{numval}
\dmsun &=&  3.7 \times 10^{-5}~{\rm eV}^2 \nn \\
\tan^2 \thetasun &=& 0.37 \quad (i.e.\;\thetasun=31^\circ )  \nn \\
\tan^2 \thetaatm &=& 0.66 \quad (i.e.\;\thetaatm=39^\circ )  \nn \\
\dmatm &=&  2.4 \times 10^{-3}~{\rm eV}^2\,.
\ea
The allowed range for $\dmlsnd$ is 1 to 0.2~eV$^2$.
However, neither the atmospheric nor the solar data are sensitive
to the exact value of $\dmlsnd$, so this parameter does not enter
our analysis.
It is instructive to compare the vacuum oscillation lengths with 
the Earth's diameter $D_\oplus$, for each of the fixed mass scales.
For the solar scale, we have 
\beq{lamsun}
\lambda_V (\dmsun)= 53\,D_\oplus\left(\frac{E}{10{\rm GeV}}\right)
	\left(\frac{3.7\times 10^{-5}~{\rm eV}^2}{\dmsun}\right)\,;
\eeq
for the atmospheric scale, we have
\beq{lamatm}
\lambda_V (\dmatm)=0.81\,D_\oplus\left(\frac{E}{10{\rm GeV}}\right)
	\left(\frac{2.4\times 10^{-3}~{\rm eV}^2}{\dmatm}\right)\,;
\eeq
and for the short-baseline scale we have
\beq{lamlsnd}
\lambda_V (\dmlsnd)=2.0\times 10^{-3}\,D_\oplus
	\left(\frac{E}{10{\rm GeV}}\right)
	\left(\frac{{\rm eV}^2}{\dmlsnd}\right)\,.
\eeq
As explained in the previous section, in earth 
these oscillation lengths are expanded near a resonance,
and become constant well above any resonance.
Nevertheless, the inferences are that the solar scale barely contributes
to atmospheric data, and then only at the lowest energies ($\lsim$~few GeV);
the SBL scale is averaged in the atmospheric data until $E\sim$~TeV;
and the atmospheric scale is well-tuned to the size of the earth
(which is why the dramatic zenith-angle dependence was so readily discovered),
and so must be handled with care.
In our numerical work this has been done.
Fast oscillations (e.g. oscillations driven by $\dmlsnd$)
are averaged over by setting large
relative phases in the oscillation probabilities to zero by hand.  
The description of this procedure is given in Appendix \ref{l/e}.

\begin{figure}
\begin{center}
\epsfxsize=120mm
\hspace*{1cm}
\epsfbox{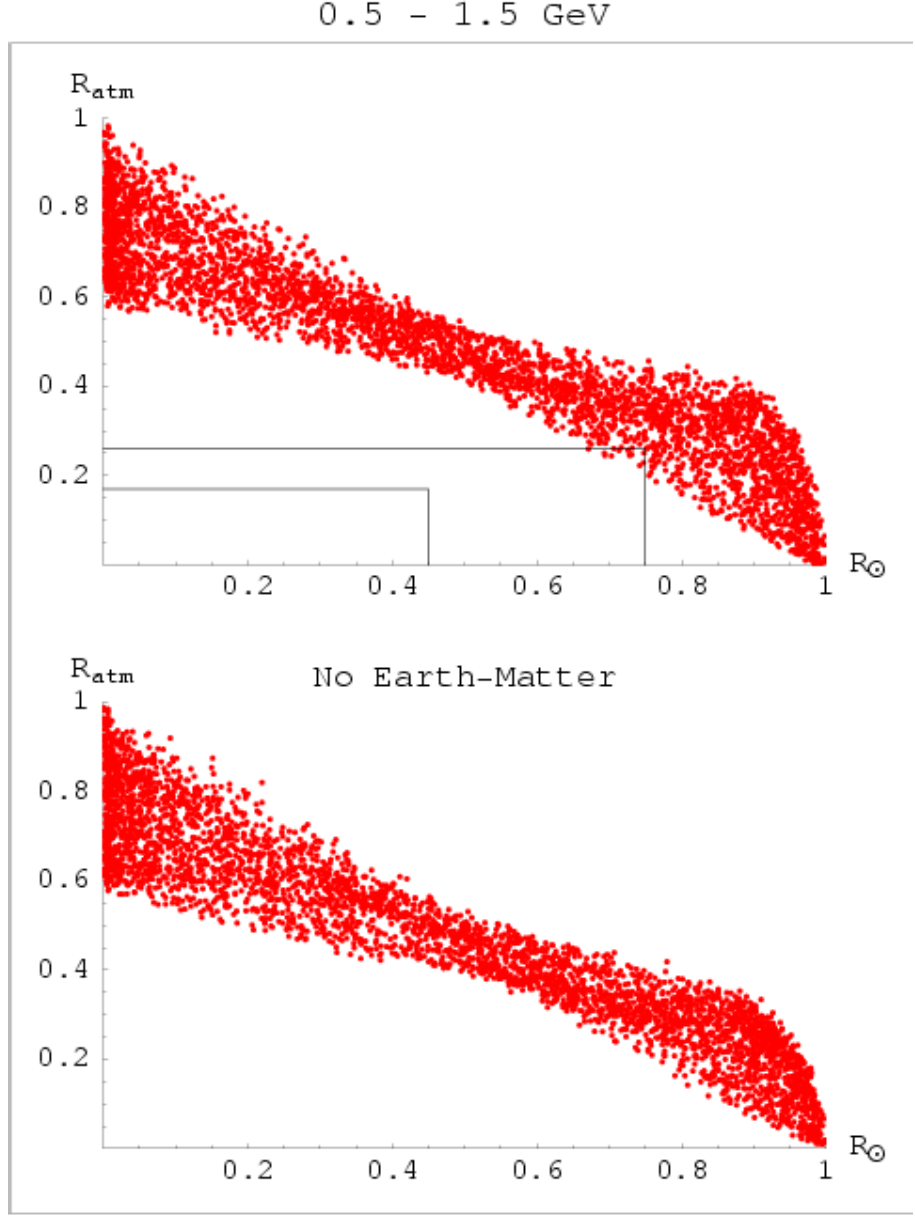}
\end{center}
\caption{Top:~4,000~points,~each~averaged over incident neutrino energies
$0.5~{\rm GeV}\le E_\nu \le 1.5~{\rm GeV}$ and upcoming angles in
$-1.0\le \cos\theta_z\le -0.8$, scattered 
over $\eps_{\mu\mu}$, 
$\eps_{\mu e}$, $\eps_{ee}$, and $\thetats$,
with matter effects included.
The 90\% and 99\% exclusion boxes obtained with small angles set to zero 
are shown as a crude reference.
Bottom: same as above but with earth-matter omitted.
In both plots, the vertical axis is $R_{\rm atm}$ and the horizontal 
axis is $R_{\rm sun}$.
\label{fig:sum_1}}
\end{figure}

\begin{figure}
\begin{center}
\epsfxsize=130mm
\hspace*{1cm}
\epsfbox{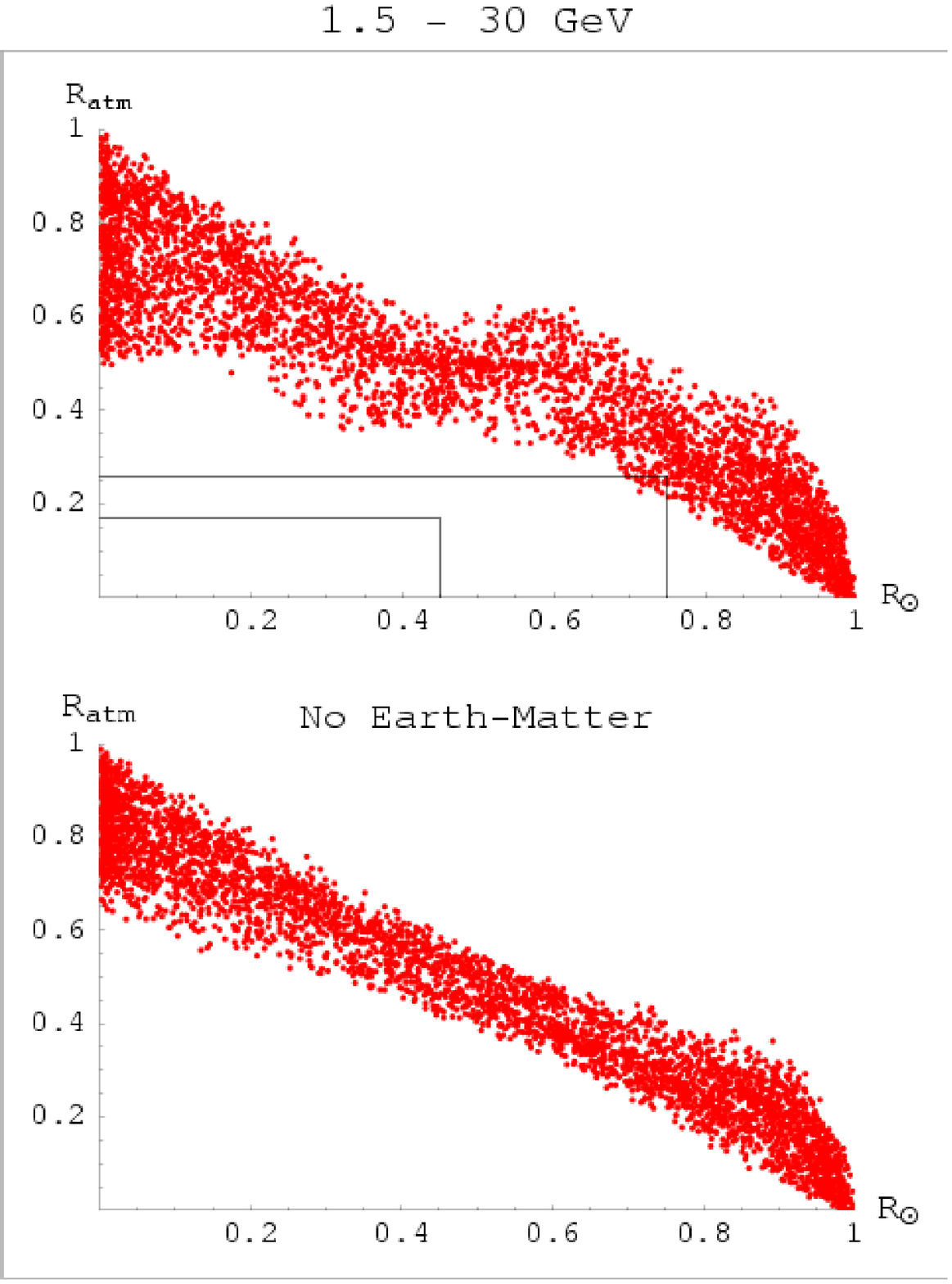}
\end{center}
\caption{Same as Fig.\ {\protect{\ref{fig:sum_1}}}, but energy-averaged over 
1.5~GeV$\le E_\nu\le 30$~GeV.
\label{fig:sum_30}}
\end{figure}

\begin{figure}
\begin{center}
\epsfxsize=130mm
\hspace*{1cm}
\epsfbox{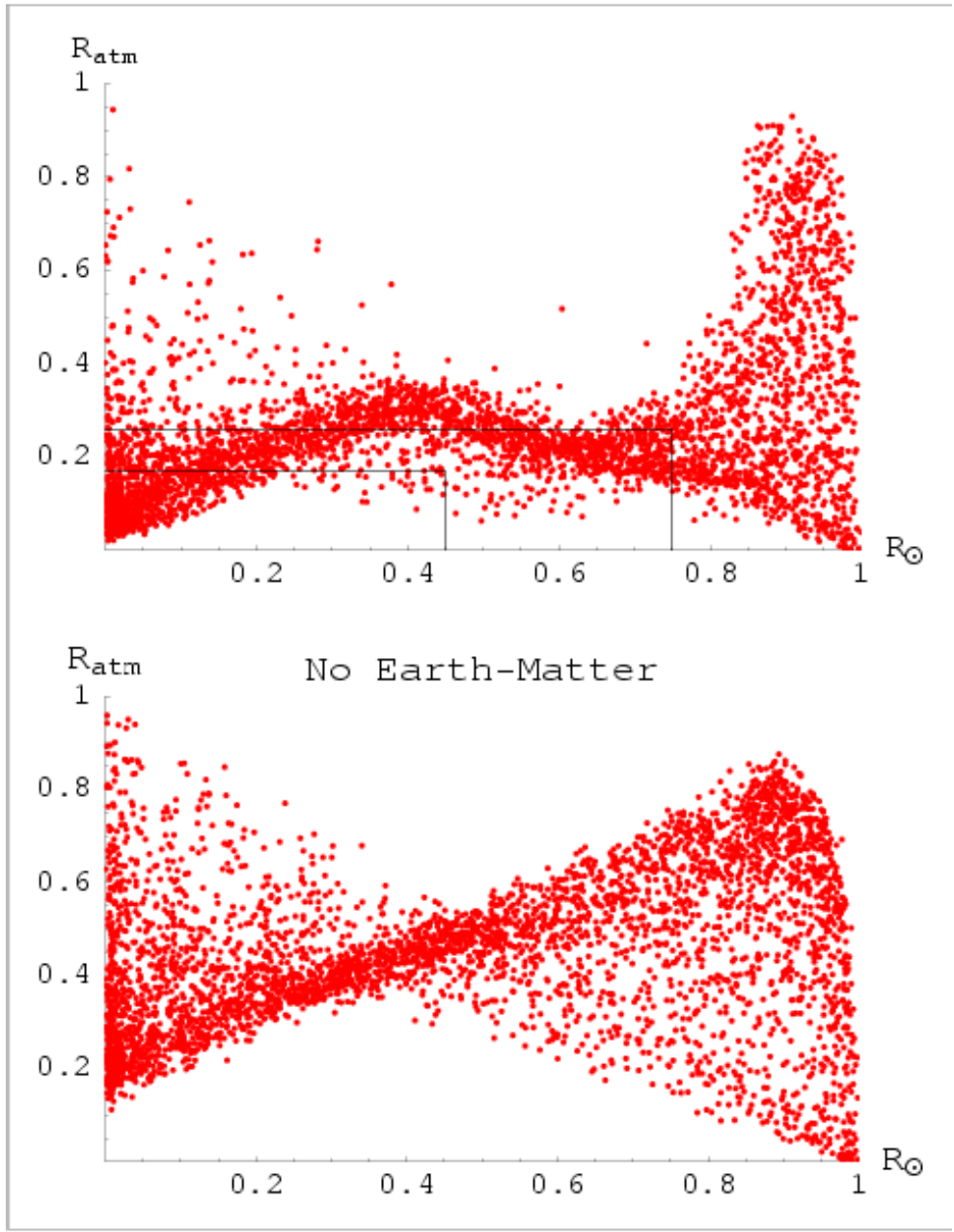}
\end{center}
\caption{Same as Fig.\ \protect{\ref{fig:sum_1}}, 
but energy-averaged over 
30~GeV$\le E_\nu\le 500$~GeV.
\label{fig:sum_500}}
\end{figure}

\begin{figure}
\begin{center}
\epsfxsize=130mm
\hspace*{1cm}
\epsfbox{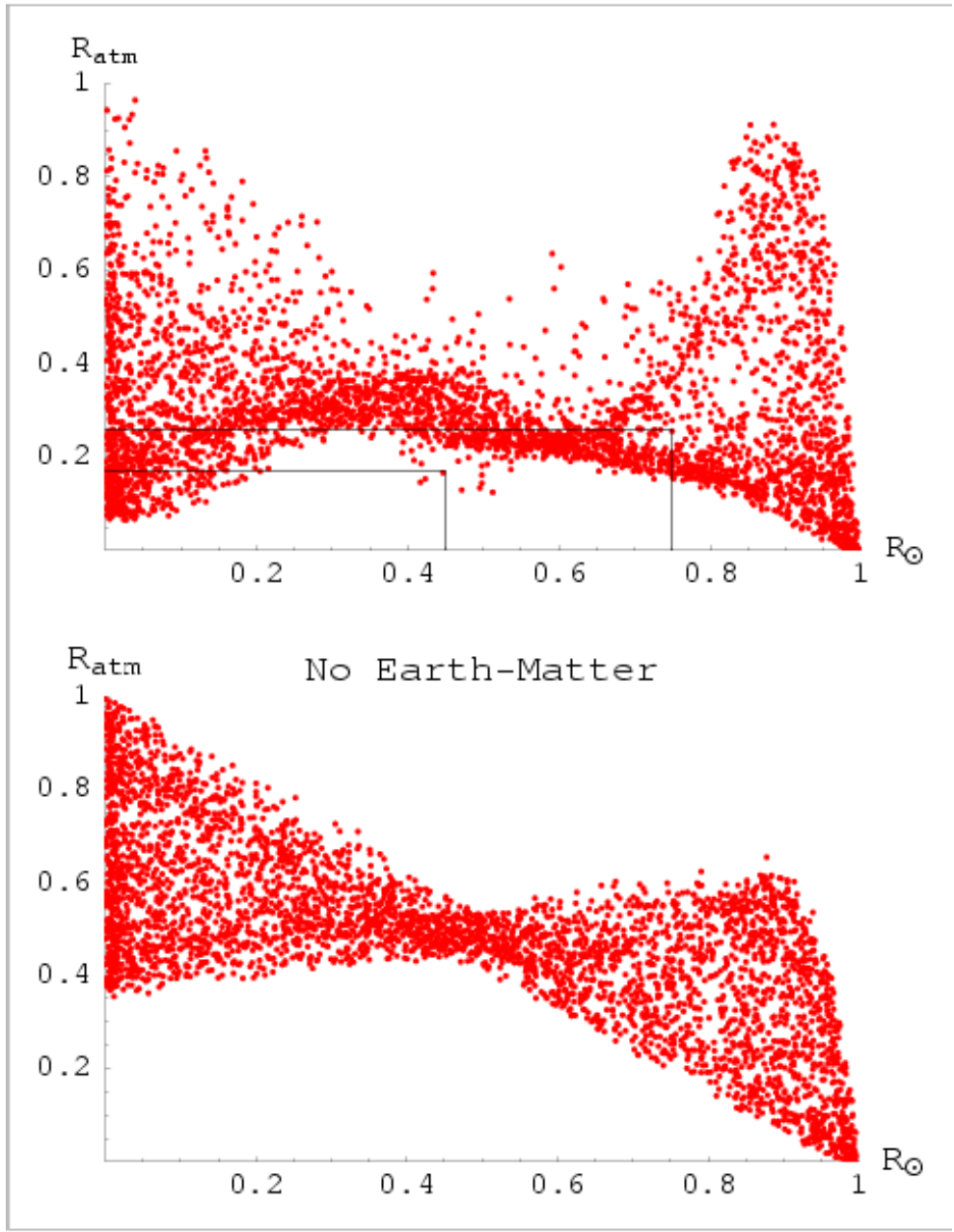}
\end{center}
\caption{Same as Fig.\ \protect{\ref{fig:sum_1}}, but energy-averaged over 
50~GeV$\le E_\nu\le 150$~GeV.
\label{fig:sum_150}}
\end{figure}

Scatter plots for the sum rule resulting from these central values
are displayed in Figs.\ \ref{fig:sum_1}-\ref{fig:sum_150}.
The number of points in each scatter plot,
and in each of those to follow, is 4,000.
This number is chosen as a balance between computer time and 
the density of the  final scatter plot.
Also shown in these figures as a guide to the eye 
are the 90\% and 99\% exclusion boxes 
that result when all three small angles are set to zero.
The weakening of the 
sum rule when small angles and matter effects are included in the 
oscillation physics is evident.
At high energies, the sum rule has so relaxed that a large 
fraction of the allowed scatter points even lie within the 
allowed region of the conservative,
zero $\eps$, exclusion boxes.
%

%
To allow the reader to assess the relative significance of 
matter-effects in earth, we show the same plots with earth matter
omitted in the bottom windows.
It is seen that earth-matter effects are not significant at low energies
but become important with increasing energy.
Near a resonance, earth-matter will affect neutrinos
and antineutinos quite differently.
Above a resonance, earth-matter will have the same effect 
on neutrinos and antineutrinos.
Whenever earth-matter effects are evident,
we have also run the same figure in the antineutrino channel 
to determine whether the matter-effect is different or the 
same for this channel.  We discuss the antineutrino channel 
in section \ref{sub:antinus}.

The earth-matter effects on the atmospheric ratio $R_{\rm atm}$ 
contributing to the sum rule \rf{SR1} are evident,
but still much less dramatic than the matter effects 
on individual atmospheric oscillation probabilities.
In particular, the sum rule has some built-in stability against 
earth-matter effects on the $\numu$ oscillation.
If the atmospheric
oscillation were pure $\numu\rightarrow\nus$, 
then the atmospheric ratio appearing in \rf{SR1} is one by definition,
independent of any matter effects.  If, at the other extreme,
the atmospheric oscillation were pure $\numu\rightarrow\nut$,
then the atmospheric ratio in \rf{SR1} is zero by definition,
again independent of any matter effects.  The intermediate cases
will show some intermediate but not large matter effect.
The ratios defined for the sum rule tend to cancel individual 
variations in 
$P(\numu\rightarrow\nus)$ and $P(\numu\rightarrow\nut)$.
In addition, the SBL contributions to the oscillation amplitudes,
arising from the large $\dmlsnd$, 
are not matter suppressed, and so tend to stabilize the sum rule.
One may expect matter effects to be more pronounced in the product 
rule \rf{SR2}, in that the stability argued here is less 
applicable for a product than for a sum, 
and less applicable for the ratios 
defined for the product rule compared to those of the sum rule.

One remarkable feature of Figs. \ref{fig:sum_1} to \ref{fig:sum_150}
is the energy dependence of the sum rule.  
A roughly diagonal band in the plots at low energy ($\lsim 30$~GeV) 
turns into a butterfly at high energy ($\gsim 30$~GeV).
This is a result of the energy-dependent matter-effects  
on the oscillation amplitude and length, and of suppression occuring
when oscillation lengths are large compared
to the Earth's diameter.

While for low energies the zero$\th$ order sum rule 
provides a good approximation to the 2+2 scheme with 
non-vanishing $\epsilon$'s, 
i.e. $R_{\rm atm} \sim \sin^2\thetats\sim 1 - R_{\rm sun}$, 
at high energies the dependence on $\theta_{\tau s}$ 
is quite different.
This is explainable.
At low energy the atmospheric oscillation is dominated by the 
LBL amplitude (\ref{numslbl}), 
leading to $\Ratm \sim \sin^2\thetats$.
However, at high energy, the LBL amplitude is
matter-suppressed at moderate to large $\thetats$ 
because then $A_{NC}$ participates
on the $\dmatm$ scale; while for
small $\thetats$, the $\numu\rightarrow\nus$ oscillations
occur dominantly via the matter-independent SBL scale, 
with amplitude given by  $4\,|\epsmm|^2\cos^2\thetats$.
We note the transition from $\sin^2\thetats$ to 
$\cos^2\thetats$ dominance as the energy grows from low to high.
It is also possible that the changing oscillation length 
plays a role in the high-energy suppression of the LBL amplitude.
As the energy increases, the wavelength grows first linearly with energy,
then reaches a maximum and finally approaches its energy-independent 
asymptotic value.
If the wavelength is long relative to the earth's diameter,
then an experiment 
samples only a small fraction of a single oscillation cycle.
For example, taking the vacuum oscillation length of 
$\lambdav (\dmatm) \sim 10\,D_\oplus\,(E_\nu/100{\rm GeV})$
as a guide, one gets a 
suppression of $\sin^2(\pi/10)=0.1$ for $E_\nu \sim 100$~GeV;
this may be compared to the SBL amplitude $4|\epsmm|^2\cos^2\thetats$, 
which may be as large as 0.48. 
To summarize this discussion, 
the atmospheric oscillation at high energy is dominated 
by the SBL contribution (\ref{SBLamp5}) proportional to $\cos^2\thetats$.
Therefore, with $\thetats \sim \pi/2$,
the sterile neutrino can hide from both the atmospheric through-going 
data and the solar data!
For through-going events ($E_\nu$ above 30~GeV),
the figures show that arbitrarily low sums seem possible.
This may explain the good fits found for the pure
$\numu\rightarrow\nus$ atmospheric model found in
\cite{Gonzalez-Garcia:2001uy,FV}.
%

One moral to be drawn here is that the SuperK through-going 
($E_\nu$ typically $\sim 100$~GeV) analysis with a 
single $\dmsq\sim 10^{-3}{\rm eV}^2$ does not apply to the 
2+2 model.  With only a LBL contribution and no SBL contribution, 
$\Ratm$ is necessarily proportional to $\sin^2\thetats$;
then non-observation of $\nus$-signatures will inevitably be
interpreted as a small value for $\thetats$ and an absence of 
$\nus$ in the atmospheric sector of the theory.


%

We have analyzed the effects on the sum rule 
of variations in the solar and atmospheric 
mixing angles, and in the three $\delta m^2$'s. 
None of these variations had dramatic effects.
The largest changes resulted from variations in 
$\dmsun$ and $\thetasun$, but especially in $\thetaatm$.
As the atmospheric angle decreases, the deviation from the sum rule 
increases.

%
%

\subsection{Individual $\eps$'s}
\label{sub:singleeps}
\begin{figure}
\begin{center}
\epsfxsize=90mm
\hspace*{1cm}
\epsfbox{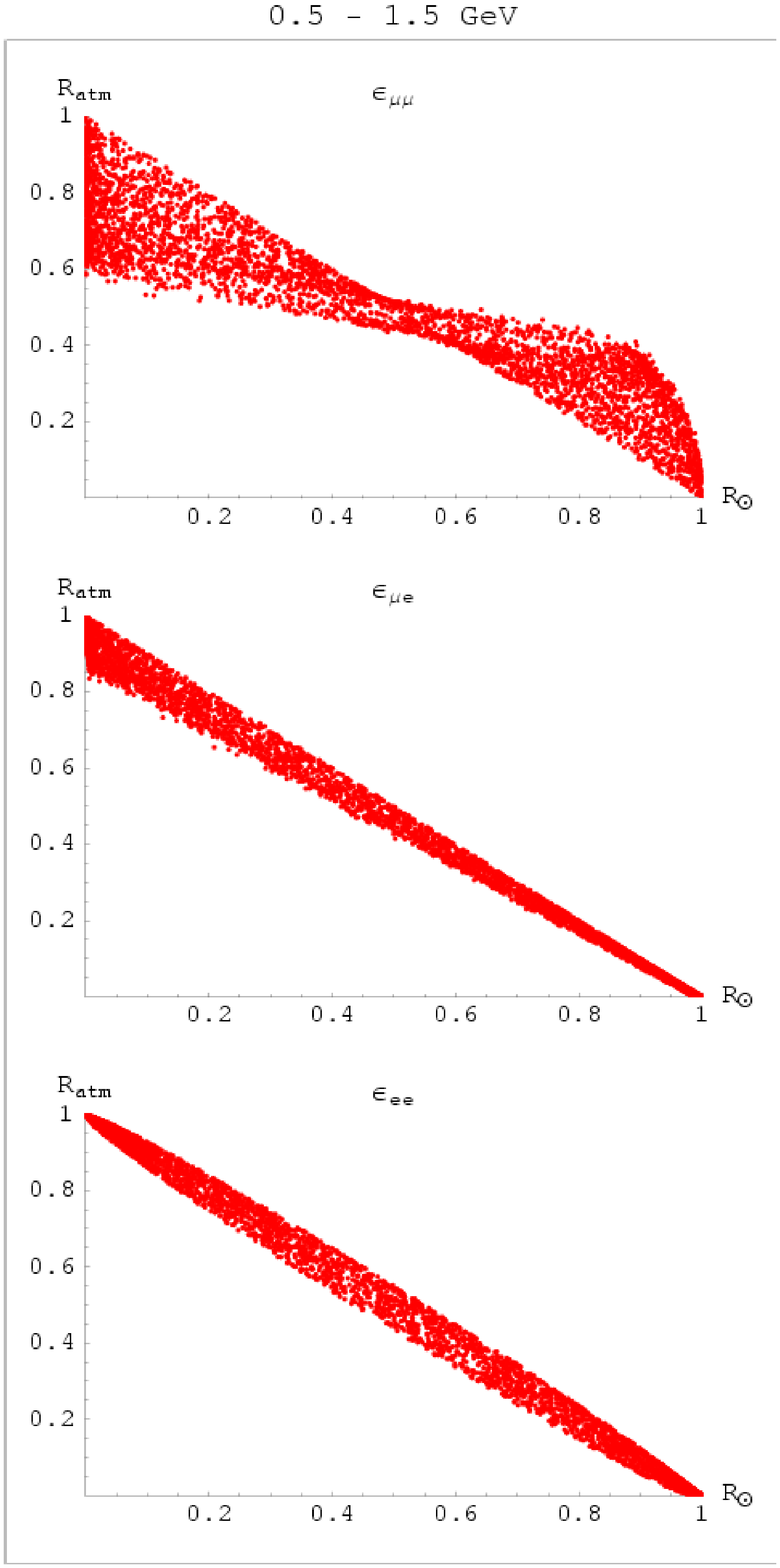}
\end{center}
\caption{Same as in Top figure \protect{\ref{fig:sum_1}},
but with only one small-angle (indicated) nonzero.
\label{fig:eps1}}
\end{figure}

\begin{figure}
\begin{center}
\epsfxsize=90mm
\hspace*{1cm}
\epsfbox{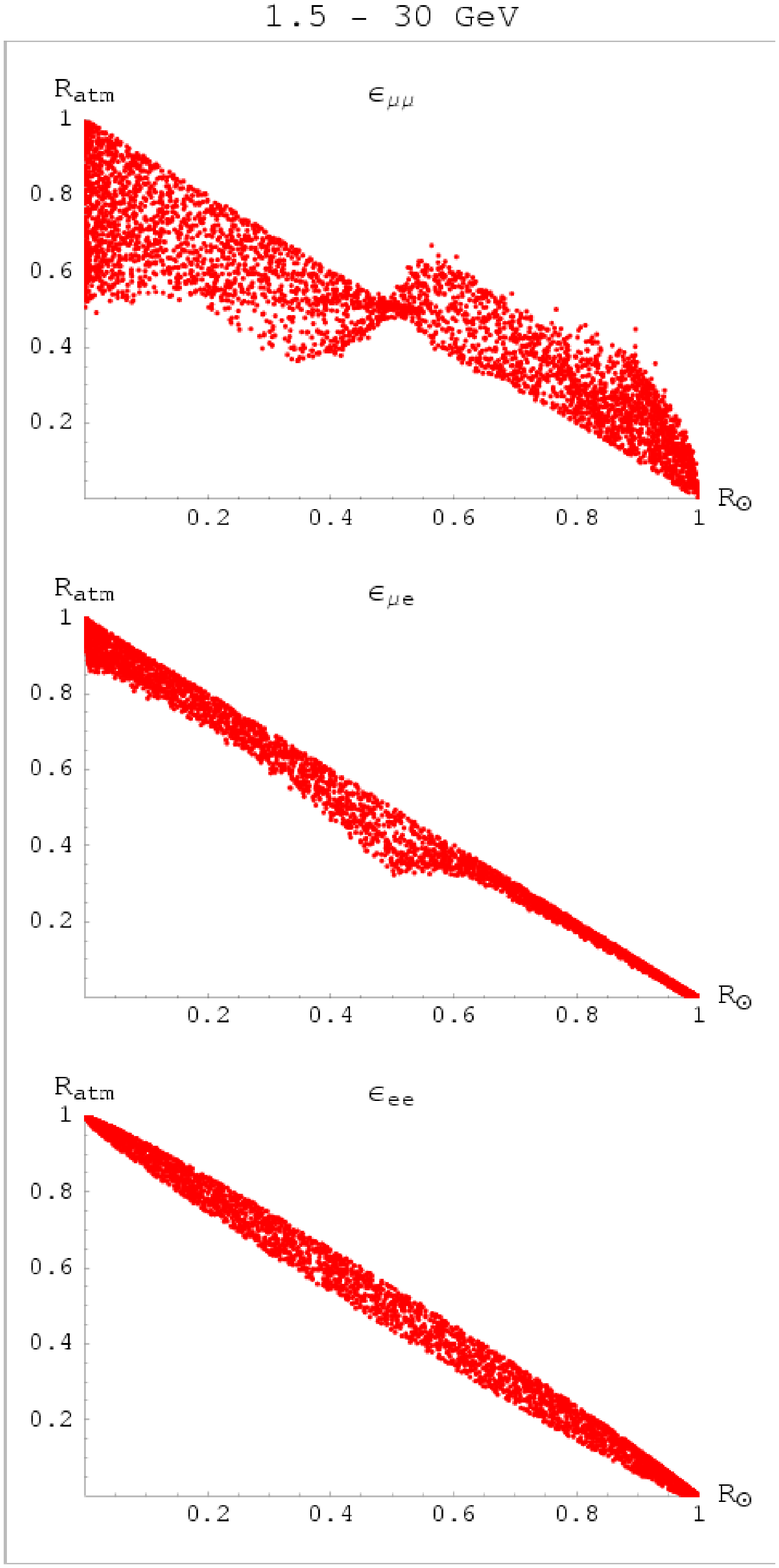}
\end{center}
\caption{Same as in Top figure \protect{\ref{fig:sum_30}},
but with only one small-angle (indicated) nonzero.
\label{fig:eps30}}
\end{figure}

\begin{figure}
\begin{center}
\epsfxsize=90mm
\hspace*{1cm}
\epsfbox{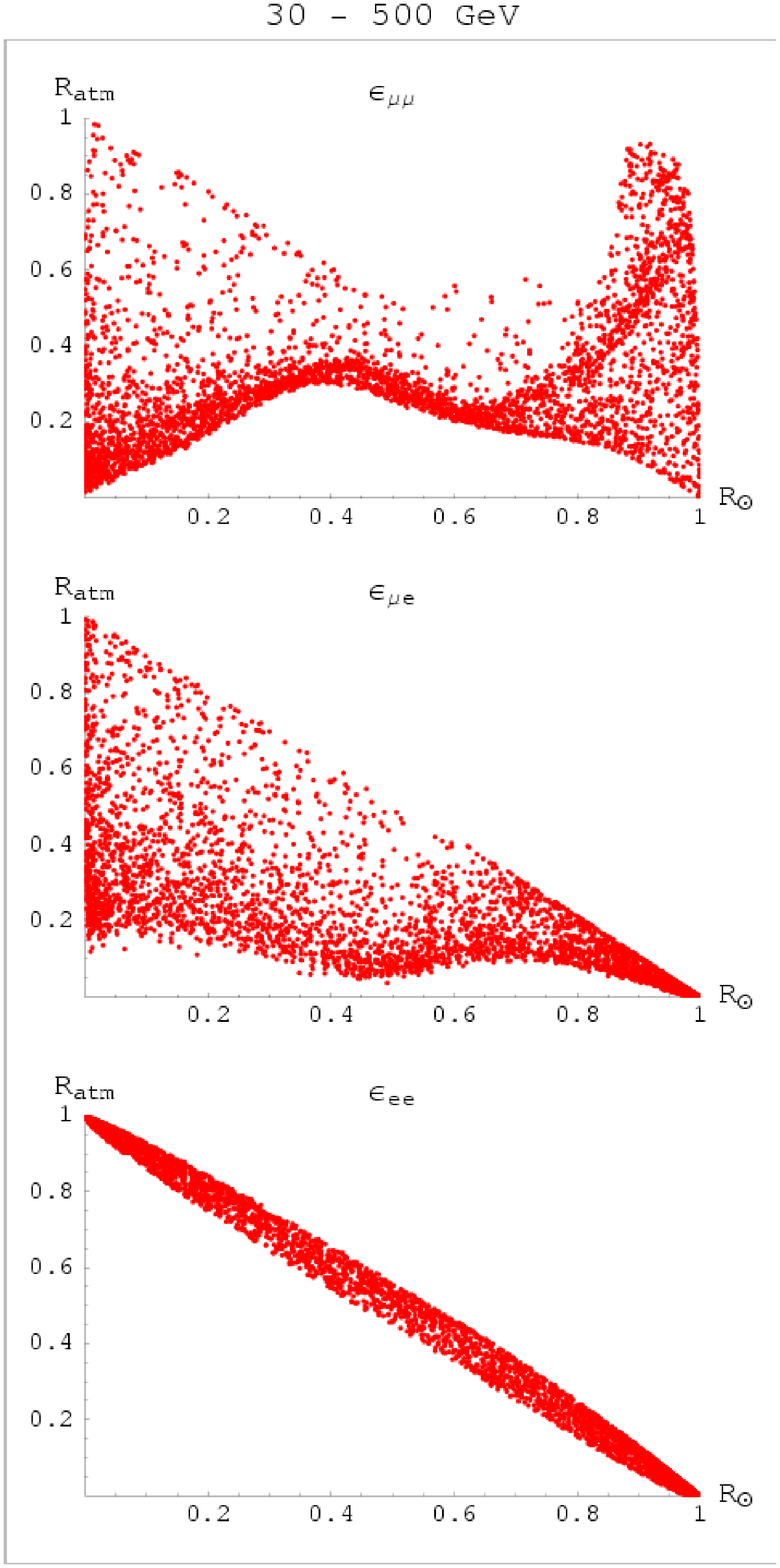}
\end{center}
\caption{Same as in Top figure \protect{\ref{fig:sum_500}},
but with only one small-angle (indicated) nonzero.
\label{fig:eps500}}
\end{figure}

\begin{figure}
\begin{center}
\epsfxsize=90mm
\hspace*{1cm}
\epsfbox{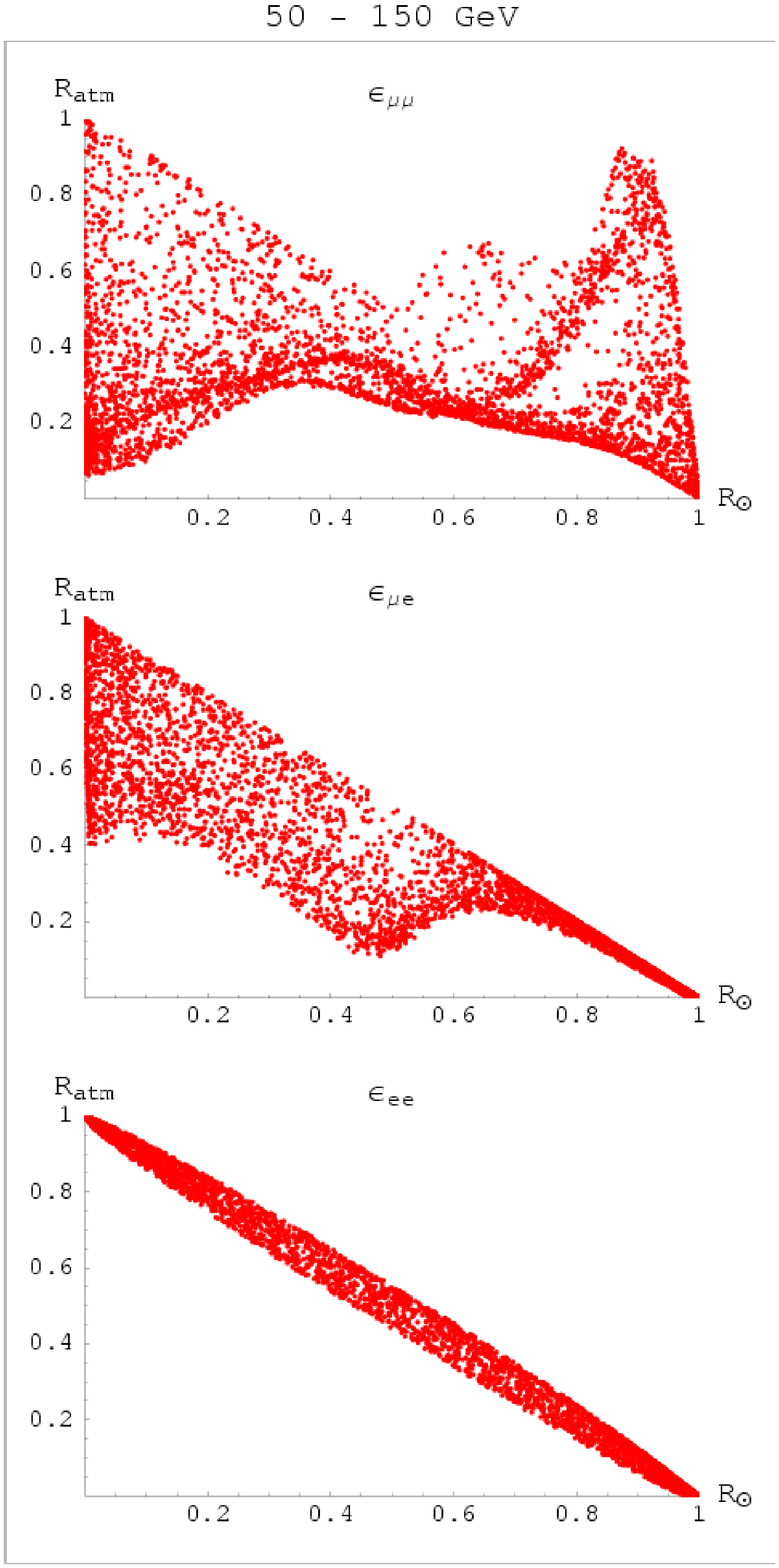}
\end{center}
\caption{Same as in Top figure \protect{\ref{fig:sum_150}},
but with only one small-angle (indicated) nonzero.
\label{fig:eps150}}
\end{figure}

An interesting question is what are the relative weights of each small 
angle in altering the sum rule result.  To answer this, 
in Figs.\ \ref{fig:eps1} to \ref{fig:eps150} we present the dependence of 
the sum rule on the individual $\eps$'s. 
A comparison shows that the effect of
non-zero $\epsee$ on the sum rule is small, and that the strong deviations 
in the sum rule at high energies are induced 
by the non-vanishing $\epsmm$ and $\epsme$. 
Especially interesting is the influence of $\epsme$, since this 
angle is set to zero in the global analyses of 
\cite{Gonzalez-Garcia:2001uy,Maltoni:2001bc,newValle}.
The results here suggest that at least the 
small mixing angles $\epsmm$ and $\epsme$ must
be consistently retained in the global analysis,
and therefore casts some doubt on recent claimed 
exclusions of the 2+2 model.

%

\subsection{Product Rule}
\label{sub:product}
\begin{figure}
\begin{center}
\epsfxsize=130mm 
\hspace*{1cm}
\epsfbox{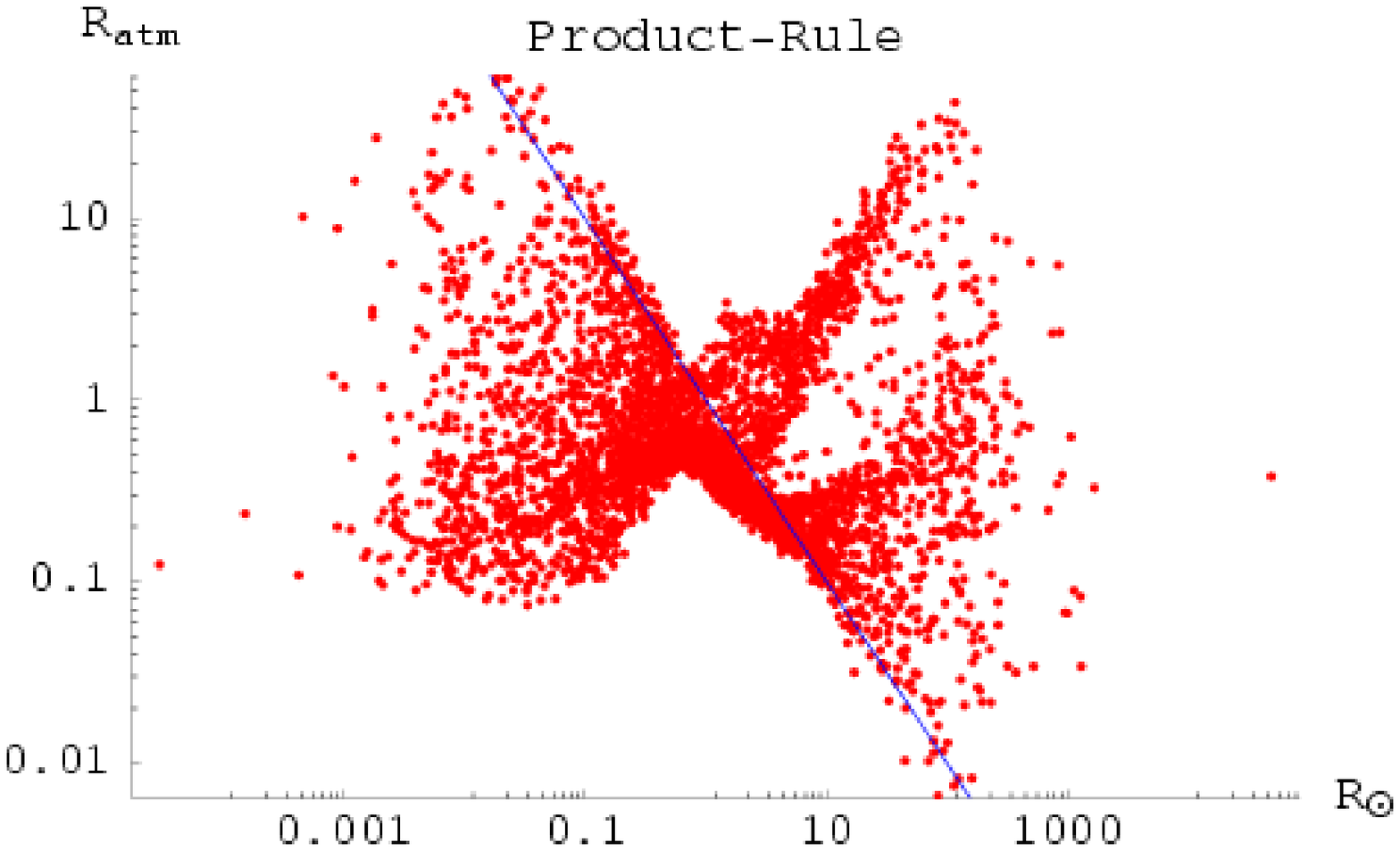}
\end{center}
\caption{4,000~points for the product rule, 
each averaged over incident neutrino energies
50~GeV~$\le E_\nu \le 150$~GeV
and upcoming angles in $-1.0\le \cos\theta_z\le -0.8$, scattered 
over $\eps_{\mu\mu}$, 
$\eps_{\mu e}$, $\eps_{ee}$, and $\thetats$, 
with matter effects included.
Here we use the same $R_{\rm atm}$ and $R_{\rm sun}$ symbols to denote the 
ratios of amplitudes appearing in the product rule eq.\ (\protect{\ref{SR2}}).
The diagonal line is the result when small angles 
are set to zero.
\label{fig:prod_150}}
\end{figure}

In Fig.\ \ref{fig:prod_150} we show the product rule for one 
selected energy range, 50~GeV~$\le E_\nu\le 150$~GeV.
Here the axes labels are 
\beq{Ratm}
R_{\rm atm}\equiv \left[
\frac{P(\numu\rightarrow\nus)}{P(\numu\rightarrow\nut)}
\right]_{\rm atm}
\eeq
and
\beq{Rsun}
R_{\rm sun}\equiv \left[
\frac{P(\nu_\odot\rightarrow\nus)}{P(\nu_\odot\rightarrow\nut)}
\right]_{\rm sun}\,.
\eeq
It is obvious that the deviation of the product rule from the 
diagonal line of the zero$\th$ order result 
is even more pronounced than the deviation of the sum rule.
We have argued in favor of this in an earlier section.

\subsection{Atmospheric Antineutrinos}
\label{sub:antinus}
Oscillation probabilities for atmospheric antineutrinos are found by 
reversing the sign of
the earth-matter potentials.
When the $\eps$'s are neglected,
matter effects on the atmospheric ratios are negligible,
and so the neutrino and antineutrino contributions to the sum rule are 
virtually
identical. With nonzero $\eps$'s, differences between
antineutrino and neutrino may appear.
Generally, we find that the antineutrino sum rule differs
little from the neutrino sum rule, even with nonzero $\eps$'s.
None of the conclusions of this paper are affected by the small
differences between atmospheric neutrino and antineutrino channels.

\begin{figure}
\begin{center}
\epsfxsize=160mm
\hspace*{1cm}
\epsfbox{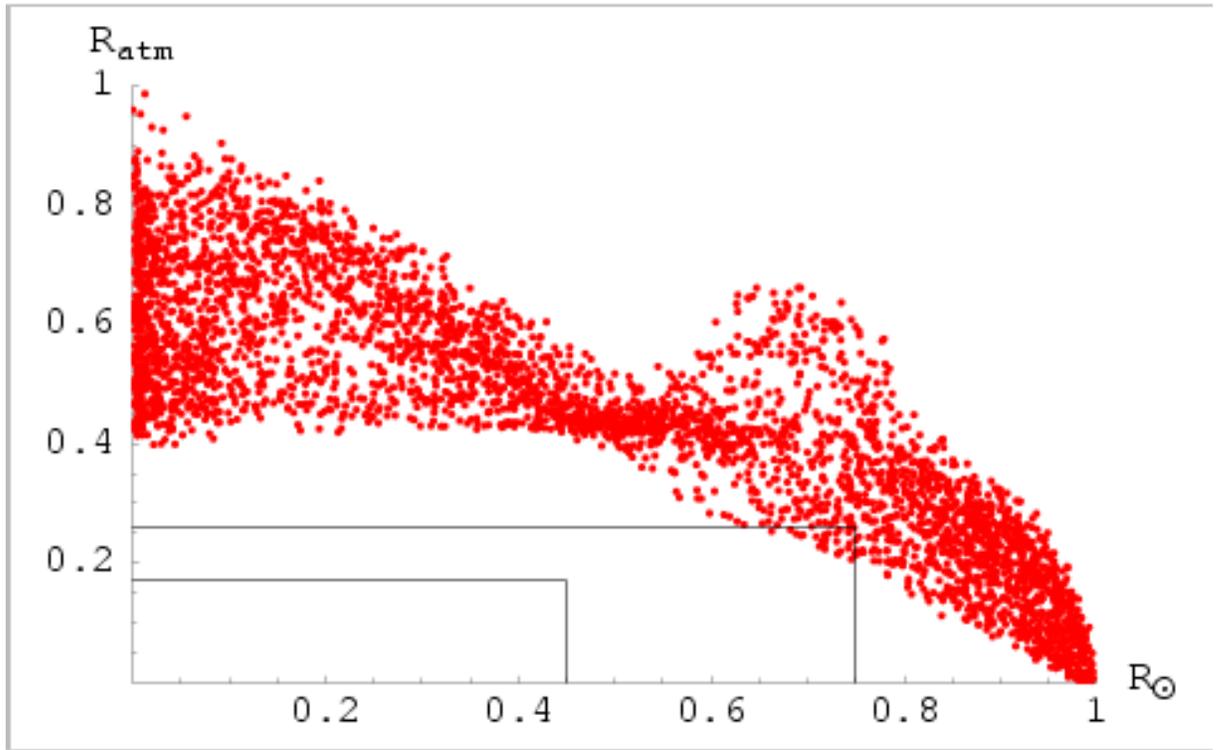}
\end{center}
\caption{Sum rule scatter plot for the antineutrino channel,
averaged over the energy range 1.5 to 30 GeV.
\label{fig:antisum}}
\end{figure}

The largest difference in the sum rule 
occurs for the energy region 1.5 to 30~GeV, which contains matter
resonances associated with the $\dmatm$~scale.
In Fig.\ \ref{fig:antisum} we show the sum rule for the antineutrino
channel in this energy region. 
This figure is to be compared to Fig.\ \ref{fig:sum_30} 
for the neutrino channel.
Larger differences between neutrino and antineutrino are found 
in the individual probabilities in this same energy region
(not shown here).
In the lower energy range 0.5 to 1.5~GeV, differences are not evident.
At high energies above the resonance region ($\gsim 30$~GeV),
matter affects both channels nearly identically.

\section{Summary and Conclusions}
In summary, we have investigated the validity of the sum 
rules \rf{SR1} and \rf{SR2} for the sterile neutrino, 
when the three small-angles in the 
general $4\times 4$ mixing matrix are included,
and when matter effects are included.
To do so, we have reduced flavor-evolution calculations for 
solar and atmospheric neutrinos to numerical 
multiplication and diagonalization of $4\times 4$ matrices.
The inputs are the vacuum mixing matrix $U_V$, the 
vacuum mass-squared (or mass-squared difference) matrix $M^2_V$,
and the electron and neutron densities for the solar core,
earth mantle, and earth core.
The output from eq.\ \rf{Hdiag} 
are the energy eigenvalues $H^\kappa$ and 
mixing matrices $U_\kappa$, $\kappa = S, M, C$, in the solar core and 
earth mantle and core, respectively.

We find that the zero$\th$ order 
sum rule offers a good approximation to the atmospheric data at low 
neutrino energies, but isn't valid at high energies.
At high energies, the long-baseline amplitude is very matter-dependent,
and the averaged short-baseline amplitude makes a significant 
contribution, even dominating for certain values of the mixing angles.
Very small values for the corrected sum rule become possible.
This means that the sterile neutrino of the 2+2 model 
can remain reasonably hidden from both solar and atmospheric 
neutrino oscillation data.
We also find that both small angles $\epsmm$ and
$\epsme$ make significant contributions to amplitudes 
when matter effects are included.
While $\epsmm$ has been included in recent combined fits to solar and 
atmospheric neutrinos \cite{Gonzalez-Garcia:2001uy,Maltoni:2001bc,newValle},
the influence of a non-vanishing $\epsme$ 
has not yet been investigated.
This weakens 
the case against the 2+2 model and the sterile 
neutrino.

We noted earlier that the important SBL amplitude, 
dominant in the atmospheric sector at high energies
for some parameter range of the 
2+2 model, is not included 
in the two angle, one $\dmsq$ fits performed by SuperK. 
This seemingly invalidates any comparison of the SuperK analysis to 
the sterile neutrino of the 2+2 model.

Related but distinct from the sum rule is a product rule
which we introduce.
The product rule experiences
even more pronounced deviations from zero$\th$ order expectations 
due to non-vanishing $\epsilon$'s and to matter.
The variables of the product rule are more difficult to extract from
data than are the variables of the sum rule.

\section{Acknowledgments}
We thank J. Conrad, A. de Gouvea, M.C. Gonzalez-Garcia, W.C. Louis, 
S. Parke, T. Schwetz, and A.Yu.\ Smirnov 
for useful discussions, and two of us (H. P. and T. J. W.) 
thank the Fermilab Summer Visitors
Program for an environment that brought about completion of this work.
This research was supported by  U. S. Department of Energy grant
number DE-FG05-85ER40226 and (HP) by the Bundesministerium f\"ur Bildung
und Forschung (BMBF, Bonn Germany) under contract number 05HT1WWA2.

\newpage

\section{Appendix A: Inclusion of Earth--Matter Effects}
In this Appendix we discuss some details of the earth-matter
effects, and their inclusion in our calculation of the sum rules.

\subsection{Atmospheric Neutrinos}
Neutrinos originating in the atmosphere traverse a chord of the earth
before measurement. 
Neutrinos which emerge from the earth at zenith angle $\theta$
greater than $33^\circ$ traverse just the mantle.
Neutrinos which emerge from the earth at zenith angle 
less than $33^\circ$ traverse the mantle, core, and more mantle,
in a fashion nearly symmetric about the midpoint.
We take the earth core to have near-constant density 
$N_e = N_n = 6\,N_A/{\rm cm}^3$ out to $R_C = 3493$~km, 
and the mantle to have near-constant density
$N_e = N_n = 1.6\,N_A/{\rm cm}^3$ out to the Earth's surface 
at $R_\oplus = 6371$~km.
When the mantle and core are approximated as constant-density layers,
it is straightforward to write down an analytic solution for flavor
propagation of atmospheric neutrinos through the earth.
 
To obtain the analytic solution,
one notes that flavor states are continuous across 
matter discontinuities, whereas mass states propagate unmixed
(i.e. diagonalize the Hamiltonian) 
within the constant-density layers. 
Thus, the analytic solution is a
product of bases-changing matrices.  
Starting with an atmospheric $\nu_\alpha$, the 
final flavor state emerging from the earth is 
\ba{DiracEarth}
|\nu_F(\theta)> =
	&|\nu_\beta>\; & <\nu_\beta|\nu_l;M>
	   <\nu_l;M|\,e^{-iH_M r_M}\,|\nu_l;M>
           <\nu_l;M|\nu_\gamma><\nu_\gamma|\nu_k;C>          \nonumber\\
        &\times\,& <\nu_k;C|\,e^{-2iH_C r_C}\,|\nu_k;C>		
                   <\nu_k;C|\nu_\sigma><\nu_\sigma|\nu_j;M>  \nonumber\\
        &\times\,& <\nu_j;M|\,e^{-iH_M r_M}\,|\nu_j;M> 
	           <\nu_j;M|\nu_\alpha>\,.
\ea
As usual, repeated indices are summed.
Inputting eqs.\ \rf{U} and \rf{Udag} into \rf{DiracEarth}, 
one gets
\beq{PsiEarth}
|\nu_F(\theta)> = U^E_{\beta\alpha}(\theta)\,|\nu_\beta>\,,
\eeq
with the unitary evolution matrix for the Earth given by
\beq{UE}
U^E (\theta) =
        \left[ U_M\,e^{-iH_M r_M}\,U_M^\dag \right]\,
        \left[ U_C\,e^{-2iH_C r_C}\,U_C^\dag \right]\,
        \left[ U_M\,e^{-iH_M r_M}\,U_M^\dag \right]\,.
\eeq
The propagation matrices $e^{-iH_\kappa r}$ are diagonal, 
with elements determined
by replacing the Hamiltonian $H_\kappa$ with its eigenvalues.
These eigenvalues, and the mixing-matrices, are obtained by 
diagonalizing (numerically) the flavor-space Hamiltonian \rf{Hflavor}.

Any matrix proportional to the identity matrix may be subtracted 
from any of the $H$'s, since this just changes the unmeasurable 
overall phase of $|\nu_F>$.

The $\theta$ dependence on the r.h.s.\ of eq.\ \rf{UE} 
is implicit in the values for the 
lengths $r_M$ and $r_C$.
For a neutrino direction with zenith angle 
$\theta < \sin^{-1} (R_C/R_\oplus) = 33^\circ$,
half of the transit distance through the core is given by
\beq{Rcore}
r_C = \sqrt{R^2_C-R^2_\oplus\,\sin^2\theta}\,,
\eeq
and 
each of the two transits through the mantle has a length given by
\beq{Rmantle}
r_M=R_\oplus\,\cos\theta - r_C\,.
\eeq
For $\theta > 33^\circ$, the path is purely in the mantle,
so $r_C=0$ and the chord length is $2 r_M=2 R_\oplus\,\cos\theta$.

For atmospheric neutrinos, we are especially interested in
the probabilities
\beq{Patm}
P_{\mu\rightarrow\beta}=|<\nu_\beta\,|U^E|\,\nu_\mu>|^2
             = |\,U^E_{\beta\mu}\,|^2
\eeq
and
\beq{}
P_{\mu\rightarrow\notmu}=1-|<\numu\,|U^E|\,\nu_\mu>|^2
             = \sum_{\beta\neq\mu}|\,U^E_{\beta\mu}\,|^2\,.
\eeq
We use this final equality as a check on our numerical work.

\subsection{Solar Neutrinos, Day-Night Differences}
For solar neutrinos arriving during the daytime, eq.\ \rf{Psun} 
applies as it is.
For neutrinos arriving at night, 
further flavor evolution (``$\nu_e$ regeneration'')
may occur in the earth-matter. The 
evolution equation \rf{PsiEarth} and its adjoint 
are used to calculate possible earth regeneration
of the $\nu_e$ component in the solar neutrino flux.
The density operator for the solar neutrino ensemble 
in the flavor basis is obtained from eq.\ \rf{rhoMass} via 
the change of basis eq.\ \rf{Udag} and its adjoint in vacuum.
Then, the flavor states are evolved according to
eq.\ \rf{PsiEarth} and its adjoint.  
The result is a density matrix for the
flavors emerging from the earth, given by
\beq{rhoF}
\rho^{S+E}_{[{\rm flavor}]} (\theta)= 
  U_E\,U_V\,D_S\,U_V^\dag\,U_E^\dag\,,
\eeq
where the weights for the density matrix are contained in 
\beq{DS}
D_S={\rm diag}\,(|U^S_{e4}|^2,\dots,|U^S_{e1}|^2)\,.
\eeq
The final result of interest is 
\beq{Pregen}
P^{S+E} (\nue\rightarrow\nu\beta)=
          (\rho^{S+E}_{[{\rm flavor}]})_{\beta\beta}\,.
\eeq
In our numerical work, 
we do not pursue earth regeneration of solar neutrinos
because the effect is known to be small.
In addition, the effect is not present in the daytime data.

\section{Appendix B: Implementing ${\bf L/E}$ resolution} \label{l/e}
Experimental reconstruction of the neutrino energy $E$ and pathlength $L$ 
is limited by information loss when the neutrino 
undergoes CC scattering, and by experimental resolution.
Such smearing in the data averages out the interferences among the
shorter-wavelength oscillations.
Also, transit of path lengths much longer than the oscillation length
leads to decoherence of the mass-eigenstate wave packets, 
again removing the interference terms.
We may implement interference loss or averaging as follows:\\
For $\theta < 33^\circ$, one has, 
from eq.\ \rf{Patm}, via eqs.\ \rf{DiracEarth} and \rf{UE}, 
\beq{phaseAveC}
P_{\mu\rightarrow\beta}=e^{-i\Phi(l,k,j;s,q,p)} 
   U_{M;\beta l}\,U_{MC;lk}\,U^\dag_{MC;kj}\,U^\dag_{M;j\mu} 
 \,U_{M;\mu p}\,U_{MC;pq}\,U^\dag_{MC;qs}\,U^\dag_{M;s\beta}\,;
\eeq
where $U_{MC}=U^\dag_M\,U_C$,
there is no sum on $\beta$ and $\mu$, 
and the phase governing the interferences is 
\ba{phaseC}
\Phi(l,k,j;s,q,p) &=& \phi(l;s)+\phi(k;q)+\phi(j;p) \nonumber \\
&=&(H^M_l-H^M_s)\,r_M+(H^C_k-H^C_q)\,(2r_C)+(H^M_j-H^M_p)\,r_M\,.
\ea
For $\theta>33^\circ$, the result is much simpler:
\beq{phaseAveM}
P_{\mu\rightarrow\beta}=e^{-i\Phi(j;p)} 
   U_{M;\beta j}\,U^\dag_{M;j\mu}\,U_{M;\mu p}\,U^\dag_{M;p\beta}\,;
\eeq
again with no sum on $\beta$ and $\mu$, 
and the phase given by 
\beq{phaseM}
\Phi(j;p)=(H^M_j - H^M_p)\,(2R_\oplus\cos\theta)\,.
\eeq

If the experimental resolution and/or reconstruction of
neutrino $L/E$ is n\%,
then a phase in excess of $\sim \frac{100}{n}\,\pi$ will be 
smeared  sufficiently that $e^{-i\Phi}$ effectively averages to zero.
Thus in numerical work,
one may set to zero contributions from values of
$\{l,k,j;s,q,p\}$ in eq.\ \rf{phaseC} 
and $\{j;p\}$ in eq.\ \rf{phaseM}
yielding a phase in excess of, say, $10\,\pi$.
In this way, all sub-dominant as well as dominant contributions
to the oscillation result are included.
In particular, phases proportional to the LSND mass-gap may be
safely set to zero.

In our work, we explicitly set the 
phases proportional to the LSND mass-gap to zero.
The other phases, proportional to $\dmatm$ or $\dmsun$,
are kept, but averaged over 50 energies and 10 zenith angles,
as explained in the main text.

\section{Appendix C: Atmospheric Density Matrix}
One may go beyond the evolution of just the single 
atmospheric flavor $\nu_\mu$.  An atmospheric density operator 
will characterize the incoherent flavor mixture produced in 
the atmosphere.
For example, taking the flavor content of neutrinos from pion decay,
one has 
\beq{rhoAtm}
{\hat \rho}^{\rm A}_{[{\rm flavor}]}= \frac{2}{3}\,|\nu_\mu><\nu_\mu|
                             +\frac{1}{3}\,|\nu_e><\nu_e| \,.
\eeq
The final density matrix is given by
\beq{rhoAtmF}
\rho^{A+E}_{[{\rm flavor}]} (\theta)= 
  U_E\,D_A\,U_E^\dag\,,
\eeq
with weight matrix $D_A=$~diag($\frac{2}{3},0,0,\frac{1}{3}$).
The final probabilities are
\beq{}
 P^{A+E}{\nu_A\rightarrow\nu_\beta}=
       \left(\rho^{A+E}_{[{\rm flavor}]}\right)_{\beta\beta}\,.
\eeq
We do not pursue this subtlety in the text.


\begin{thebibliography}{99}

\bibitem{miniBooNE} miniBooNE homepage: http://www-boone.fnal.gov

\bibitem{BGG} 
S.~M.~Bilenkii, C.~Giunti and W.~Grimus,
Eur.\ Phys.\ J.\ C {\bf 1}, 247 (1998)
[arXiv:hep-ph/9607372].

\bibitem{Barger:1998bn}
V.~Barger, S.~Pakvasa, T.~J.~Weiler and K.~Whisnant,
Phys.\ Rev.\ D {\bf 58}, 093016 (1998)
[hep-ph/9806328].

\bibitem{Mills:2001tq}
G.~B.~Mills  [LSND Collaboration],
Nucl.\ Phys.\ Proc.\ Suppl.\ {\bf 91}, 198 (2001).

\bibitem{Barger:2000ch}
V.~Barger, B.~Kayser, J.~Learned, T.~Weiler and K.~Whisnant,
Phys.\ Lett.\ B {\bf 489}, 345 (2000)
[hep-ph/0008019].

\bibitem{Smirnov:2000pa}
A.~Y.~Smirnov,
Nucl.\ Phys.\ Proc.\ Suppl.\ {\bf 91}, 306 (2000)
[hep-ph/0010097].


\bibitem{Giunti:2001ur}
C.~Giunti and M.~Laveder,
JHEP{\bf 0102}, 001 (2001)
[hep-ph/0010009].

\bibitem{Grimus:2001mn}
W.~Grimus and T.~Schwetz,
Eur.\ Phys.\ J.\ C {\bf 20}, 1 (2001)
[arXiv:hep-ph/0102252].

\bibitem{Peres:2001ic}
O.~L.~Peres and A.~Y.~Smirnov,
Nucl.\ Phys.\ B {\bf 599}, 3 (2001)
[hep-ph/0011054].

\bibitem{newSKlimit} 
A.~Habig  [SuperKamiokande Collaboration],
arXiv:hep-ex/0106025;
M. Shiozawa (SuperK Collaboration), 
presented at ``Neutrino2002,'' Munich, Germany, May 2002.

\bibitem{FLM01}
G. Fogli, E. Lisi, and A. Marrone, Phys. Rev. D63, 053008 (2001).


\bibitem{Giunti:2000wt}
C.~Giunti, M.~C.~Gonzalez-Garcia and C.~Pe\~{n}a-Garay,
Phys.\ Rev.\ D {\bf 62}, 013005 (2000)
[arXiv:hep-ph/0001101].



\bibitem{BMW01}
V.~D.~Barger, D.~Marfatia and K.~Whisnant,
arXiv:hep-ph/0106207.

\bibitem{snosterile}
J.~N.~Bahcall, M.~C.~Gonzalez-Garcia and C.~Pe\~{n}a-Garay,
arXiv:hep-ph/0204314,
V.~Barger, D.~Marfatia, K.~Whisnant and B.~P.~Wood,
arXiv:hep-ph/0204253,
J.~N.~Bahcall, M.~C.~Gonzalez-Garcia and C.~Pe\~{n}a-Garay,
arXiv:hep-ph/0204194;
A. Bandyopadhyay, S. Choubey, S. Goswami, and D.P. Roy,
Phys. Lett. B540, 14 (2002).

\bibitem{Gonzalez-Garcia:2001uy}
M.~C.~Gonzalez-Garcia, M.~Maltoni and C.~Pe\~{n}a-Garay,
Phys.\ Rev.\ D {\bf 64}, 093001 (2001)
[arXiv:hep-ph/0105269].
M.~C.~Gonzalez-Garcia, M.~Maltoni and C.~Pe\~{n}a-Garay,
arXiv:hep-ph/0108073.


\bibitem{Maltoni:2001bc}
M.~Maltoni, T.~Schwetz and J.~W.~Valle,
arXiv:hep-ph/0112103.

\bibitem{newValle}
M.~Maltoni, T.~Schwetz, M.~A.~Tortola, and J.~W.~Valle,
arXiv:hep-ph/0207157 and hep-ph/0207227.

\bibitem{Gronau:1985kx}
M.~Gronau, R.~Johnson and J.~Schechter,
Phys.\ Rev.\ D {\bf 32}, 3062 (1985).


\bibitem{Dooling00}
D.~Dooling, C.~Giunti, K.~Kang and C.~W.~Kim,
Phys.\ Rev.\ D {\bf 61}, 073011 (2000)
[arXiv:hep-ph/9908513];
also, \cite{FLM01} for the zero $\eps$ analysis;
these works extends the much earlier three-neutrino work of 
T. K. Kuo and J. Pantaleone, Phy. Rev. D35, 3432 (1987).

\bibitem{SNOwebsite}
http://www.sno.phy.queensu.ca/

\bibitem{Declais:1994su}
Y.~Declais {\it et al.},
Nucl.\ Phys.\ B {\bf 434}, 503 (1995).

\bibitem{Dydak:1983zq}
F.~Dydak {\it et al.},
Phys.\ Lett.\ B {\bf 134}, 281 (1984).

\bibitem{future} H. P\"as, L. Song, and T.J. Weiler, in preparation.

\bibitem{snolma}
P.~C.~de Holanda and A.~Y.~Smirnov,
arXiv:hep-ph/0205241,
A.~Strumia, C.~Cattadori, N.~Ferrari and F.~Vissani,
arXiv:hep-ph/0205261,
A.~Bandyopadhyay, S.~Choubey, S.~Goswami and D.~P.~Roy,
arXiv:hep-ph/0204286.

\bibitem{FV} R. Foot, Phys. Lett. B496, 169 (2000);
ibid, arXiv:hep-ph/0210393;
R. Foot and R. Volkas, Phys. Lett. B543, 38 (2002).


\end{thebibliography}


\newpage

\newpage

\newpage

\newpage

\newpage

\newpage

\newpage

\newpage

\newpage








\newpage

\newpage




\end{document}